\documentclass{article}

\usepackage{arxiv}
\renewcommand{\headeright}{}
\renewcommand{\undertitle}{}
\renewcommand{\shorttitle}{}

\usepackage[utf8]{inputenc}
\usepackage[T1]{fontenc}
\usepackage{microtype}

\usepackage{booktabs}
\usepackage{subcaption}

\usepackage{amsmath}
\usepackage{amssymb}
\usepackage{amsfonts}
\usepackage{mathtools}
\usepackage{stmaryrd}
\usepackage{physics}
\usepackage{nicematrix}
\DeclareMathOperator{\diag}{diag}
\DeclareMathOperator{\poly}{poly}

\usepackage{zx-calculus}

\usepackage{amsthm}
\usepackage{thmtools}
\usepackage{thm-restate}
\declaretheoremstyle[
spaceabove=10pt plus 2.0pt minus 4.0pt,
spacebelow=10pt plus 2.0pt minus 4.0pt,
headfont=\normalfont\bfseries,
notefont=\normalfont\bfseries,
bodyfont=\normalfont,
postheadspace=\newline,
headpunct=.{},
qed=\ensuremath{\blacksquare}
]{theoremstyle}

\declaretheorem[style=theoremstyle,parent=section]{proposition}
\declaretheorem[style=theoremstyle,sibling=proposition]{definition}

\usepackage{cite}
\usepackage[
colorlinks,
citecolor=black,
linkcolor=black]{hyperref}
\usepackage[nameinlink]{cleveref}

\title{Double-Logarithmic Depth Block-Encodings \\ of Simple Finite Difference Method's Matrices}
\date{\today}
\author
{
{\hspace{-0.75mm}Sunheang TY} \\
IRT SystemX \\
Gif-sur-Yvette, France \\
\texttt{sunheang.ty@irt-systemx.fr} \\
\And
{\hspace{-0.75mm}Renaud Vilmart} \\
Université Paris-Saclay \\
ENS Paris-Saclay, CNRS, Inria, LMF \\
Gif-sur-Yvette, France \\
\texttt{renaud.vilmart@inria.fr} \\
\And
{\hspace{-7.85mm}Axel TahmasebiMoradi} \\
{\hspace{-7.1mm}IRT SystemX} \\
{\hspace{-7.1mm}Gif-sur-Yvette, France} \\
{\hspace{-7.1mm}\texttt{a.tahmasebimoradi@irt-systemx.fr}} \\
\And
{\hspace{-5.85mm}Chetra Mang} \\
{\hspace{-5.1mm}IRT SystemX} \\
{\hspace{-5.1mm}Gif-sur-Yvette, France} \\
{\hspace{-5.1mm}\texttt{chetra.mang@irt-systemx.fr}} \\
}

\begin{document}

\raggedbottom
\allowdisplaybreaks
\setlength{\intextsep}{15pt plus 2.0pt minus 4.0pt}
\setlength{\textfloatsep}{15pt plus 2.0pt minus 4.0pt}
\setlength{\abovedisplayskip}{7.5pt plus 2.0pt minus 4.0pt}
\setlength{\belowdisplayskip}{7.5pt plus 2.0pt minus 4.0pt}

\maketitle

\begin{abstract}
Solving differential equations is one of the most computationally expensive problems in classical computing, occupying the vast majority of high-performance computing resources devoted towards practical applications in various fields of science and engineering. Despite recent progress made in the field of quantum computing and quantum algorithms, its end-to-end application towards practical realization still remains unattainable. In this article, we tackle one of the primary obstacles towards this ultimate objective, specifically the encoding of matrices derived via finite difference method solving Poisson partial differential equations in simple boundary-value problems. To that end, we propose a novel methodology called \emph{block-diagonalization}, which provides a common decomposition form for our matrices, and similarly a common procedure for block-encoding these matrices inside a unitary operator of a quantum circuit. The depth of these circuits is double-logarithmic in the matrix size, which is an exponential improvement over existing quantum methods and a superexponential improvement over existing classical methods. These improvements come at the price of a constant multiplicative overhead on the number of qubits and the number of gates. Combined with quantum linear solver algorithms, we can utilize these quantum circuits to produce a quantum state representation of the solution to the Poisson partial differential equations and their boundary-value problems.
\end{abstract}


\section{Introduction}
\label{section : introduction}

Differential equations are ubiquitous in the vast majority of science and engineering fields, with an extensive number of practical applications such that a significant research-and-development in high-performance computing field are devoted towards solving these problems. Unsurprisingly, there have been several proposals for the uses of quantum computing to solve such important problems. There are several different categories and classes of differential equations depending on the fields and applications of the problem. In this article, we are interested in linear partial differential equations \cite{evans2022partial}, which constitute the majority of practical applications. More specifically, we are interested in the boundary-value problems comprising a second-order linear partial differential equation called Poisson's equation within hypercube domains and a number of different boundary conditions. And although, such problems might be solved via analytical method, we are interested in the solution via a numerical method called finite difference \cite{smith1985numerical,leveque2007finite,thomas2013numerical}, as it can serve as the foundation for other numerical methods such as finite element method and finite volume method. Basically, finite difference method transforms the boundary-value problem, a continuous problem, into a system of linear equations, a discrete problem, by discretizing the continuous domain into a collection of grid points, and approximating the partial derivative of the solution locally at each grid point. The derived system of linear equations is generally sparse and can be solved using classical linear solver algorithms \cite{saad2003iterative,vishnoi2013lx}, such as conjugate gradient method. Its time complexity is linear in the number of grid points or the matrix size, and logarithmic in the inverse of solution error. Note that the number of grid points, and therefore the matrix size, is related to the discretization error introduced by the method; thus, the time complexity is dependent on both errors. The solution from the linear solver is the discretized solution of the differential equation at each grid point. Generally, for practical applications, the number of grid points required are significant, such that high-performance computing resources are required; thus, providing the motivation for further research.

Recently, with the advances of quantum computing \cite{nielsen2010quantum}, and quantum algorithms \cite{dalzell2023quantum} in particular, there have been a surge of research-and-development focusing on the application of these advances towards some of the most computationally expensive problems in classical computing. These include problems in the field of condensed matter physics, nuclear and particle physics, quantum chemistry, differential equations, combinatorial optimization, continuous optimization, and machine learning. Solving a single problem in a particular field generally requires a combination of several quantum algorithmic primitives to compose an overall quantum algorithm, and incorporate into it quantum error correction and fault tolerance in order to correctly implement it on a quantum computer. In this article, we are interested in the quantum algorithmic primitives for solving differential equations, in particular those by means of quantum linear solver algorithms \cite{harrow2009quantum,childs2017quantum,gilyen2019quantum,martyn2021grand,tong2021fast,algassertUsingQuantum,subacsi2019quantum,an2022quantum,lin2020optimal,costa2022optimal,jennings2023efficient,huang2019near,bravo2019variational,xu2021variational}, i.e., those that transformed differential equations into a system of linear equations, such as finite difference method. These algorithms \cite{harrow2009quantum,childs2017quantum,gilyen2019quantum,martyn2021grand,tong2021fast,algassertUsingQuantum,subacsi2019quantum,an2022quantum,lin2020optimal,costa2022optimal,jennings2023efficient,huang2019near,bravo2019variational,xu2021variational} have been improved over the years using various different techniques, from quantum phase estimation to quantum singular value transformation. Currently, the best known quantum linear solver is \cite{costa2022optimal} using the adiabatic theorem, whose time complexity is linear in the matrix's condition number and logarithmic in the inverse of solution error. Denote $\mathcal{L}\ket{u}=\ket{f}$ as the system of linear equations. It requires two quantum primitive inputs: a block-encoding quantum circuit of $\mathcal{L}$, and a quantum state preparation of $\ket{f}$, and it produces a quantum state representation of $\ket{u}$ as an output. To take full advantage of this algorithm, five efficient quantum algorithmic primitives are required \cite{aaronson2015read}: a quantum linear solver itself, a quantum preconditioner, an encoding of $\mathcal{L}$, an encoding of $\ket{f}$, and a decoding of quantum state representation of $\ket{u}$. Each of these five primitives presents its own challenges, and can depend upon the intended applications. In this article, we are tackling the third obstacle: the encoding of matrices derived via finite difference method solving Poisson partial differential equations in boundary-value problems, with the hope that it can play an important role in the ultimate objective of an end-to-end application of quantum computing.

\paragraph{Main Contributions}

The main contribution of this article is the \emph{block-diagonalization} methodology, which describes a common decomposition form for the matrices derived by the discretization via finite difference method of boundary-value problems described by a Poisson partial differential equation in a hypercube domain of an arbitrary dimension, and four different boundary conditions: periodic, Dirichlet, Neumann, and Robin, which are typical study cases in both academic and industry. The primary benefit of this common decomposition form is that it also leads to a common procedure for the construction of quantum circuits block-encoding each of these matrices inside a unitary operator. Thus, avoiding the need to construct a quantum oracle which can compute the entries of the matrices given its row and column position; such oracle is much harder to construct. Another main benefit of block-diagonalization is that their block-encoding quantum circuits are simple. In fact, they are simple enough such that we are able to derive their simplifications directly without using any sophisticated circuit optimization software. Moreover, the simplified quantum circuits comprise only a few ancilla qubit, elementary quantum gates and circuits. Using existing quantum gate and circuit implementation techniques, especially the families of classical reversible circuits including arithmetic circuits, we are able to show that these elementary gates and circuits can be implemented in logarithmic depth in the number of qubit. These come at the price of a constant multiplicative overhead on the number of qubits and gates. Therefore, the implementations of our block-encoding quantum circuits also have double-logarithmic depth in the matrix size.

Unfortunately, existing quantum methods, such as the aforementioned quantum oracle, have polylogarithmic depth, while existing classical methods have linear depth. Combine these with quantum and classical linear solver algorithms, we find our method's dependencies to be exponentially better than existing quantum methods, and superexponentially better than existing classical methods in terms of matrix size, for solving our boundary-value problems. However, note that the matrix size has a dependency in the discretization error. Assuming this error is of the same order as the solution error of the quantum linear solver, our method is still polynomially and exponentially better than existing quantum and classical methods, respectively. These improvements are based on the assumption that the quantum state preparation subroutine has at most the same complexity as our block-encoding. Another caveat is that like all quantum methods, we keep the solution as a quantum state, rather than a classical state; otherwise, most of the quantum speedup are lost.

\paragraph{Content Outline}

This article is divided into six sections. \Cref{section : introduction} introduces the state-of-the-art in the uses of quantum computing for solving differential equations, by presenting the advantages of quantum computing, and identifying the challenges of its application to this problem domain. It also describes the main contribution of the article, outlines its structure, and reviews related works. \Cref{section : preliminaries} provides background information about the boundary-value problems discussed in this article. It comprises the Poisson partial differential equation, the four boundary conditions, i.e., periodic, Dirichlet, Neumann, and Robin, and their extensions to higher dimensional problems. For each problem, its discretization as a system of linear equations via finite difference method is presented. \Cref{section : block-diagonalization} presents the main contribution of the article. It begins with a definition of a \emph{block-diagonalizable} matrix, and shows that the discretized boundary-value problems' matrices presented in the preceding section are all block-diagonalizable, i.e., they can all be put into a common form. \Cref{section : block-encoding} takes advantages of this common form, to construct quantum circuits which can block-encode each of the aforementioned matrices inside a unitary operator. For each quantum circuit, a decomposition into elementary quantum gates and circuits is also provided. \Cref{section : analyses} analyses the computational resources required to construct such block-encoding quantum circuits, and the time complexity of their applications with quantum linear solver algorithms to solve our boundary-value problems. \Cref{section : discussion} completes the article by reiterating the main contribution of the article, discussing its overall impact on the quantum computing domain for solving differential equations and beyond, and outlining some future research directions regarding the block-diagonalization methodology.

\paragraph{Related Works}

Since the inception of the first quantum linear solver algorithm, there have been several proposals regarding their application towards finite difference method and other numerical methods in general. \cite{cao2013quantum,childs2021high,vazquez2022enhancing,kharazi2024explicit} are some proposals for finite difference method and spectral method solving Poisson equations, using various quantum linear solver algorithms, discretization schemes, and matrix encoding techniques. \cite{clader2013preconditioned} proposes a quantum preconditioner and demonstrates its application for finite element method, while \cite{montanaro2016quantum} provides the first end-to-end time complexity analysis by including the dependencies between the matrix size and the discretization error for finite element method. Other works focus on the differential equations themselves, including wave equation \cite{costa2019quantum}, heat equation \cite{linden2022quantum}, linear and non-linear ordinary differential equations \cite{leyton2008quantum,berry2014high,berry2017quantum,childs2020quantum,liu2021efficient,an2022theory,krovi2023improved,liu2023efficient}, and non-linear partial differential equations \cite{gaitan2020finding,gaitan2021finding,oz2022solving}. And finally, there are also works related to the block-encoding of both dense and sparse matrices with special structures \cite{clader2022quantum,nguyen2022block,camps2024explicit,sunderhauf2024block,li2023efficient}.

In this work, we are particularly interested in works related to the encoding of finite difference method's matrices of our boundary-value problems using (1) the conventional quantum oracles \cite[superexponential]{dalzell2023quantum}, capable of computing the entries of the matrix given their row and column indices, (2) sparse matrices derived by adaptive central-difference approximation scheme \cite{childs2021high}, and (3) sparse matrices derived by 3-points central-difference approximation scheme \cite
{kharazi2024explicit}, similar to our matrices in this article. For each of these encoding techniques, we provide analyses of these encoding techniques, using either the prescribed quantum linear solver algorithm provided by each of the article, or the quantum linear solver from \cite{costa2022optimal} which is used in this article, to ensure fair comparisons. Other encoding techniques of finite difference method matrices include \cite{cao2013quantum,vazquez2022enhancing}; however, they can be subsumed by \cite{childs2021high,kharazi2024explicit} due to their similarities.

\section{Preliminaries}
\label{section : preliminaries}

A boundary-value problem comprises two components: a differential equation and a boundary condition. In this article, we study a typical second-order linear partial differential equation called Poisson's equation as given in \cref{equation : Poisson's equation}, and a periodic boundary condition as given in \cref{equation : periodic boundary condition}, where $u,f : [0,1] \rightarrow \mathbb{R}$ are solution and data function, respectively.
\begin{align}
-\frac{d^2}{dx^2}
u(x)
&=
f(x)
\quad
\text{for all}
\quad
x \in (0,1)
\label{equation : Poisson's equation}
\\
u(0)
&=
u(1)
\label{equation : periodic boundary condition}
\end{align}

Such problem can be solved numerically by finite difference method, whereby the solutions are the discretized values of $u$ at some particular grid points. In this article, we use uniform discretization of $N$ grid points and define $h := 1/N$ as the grid size. Then, using a $3$-points central-difference approximation scheme, we derive a system of linear equations \cref{equation : system of linear equations : periodic boundary condition}, where $(u_j, f_j) := (u(jh), f(jh))$ for all $j \in \{0, \ldots, N-1\}$, are the discretized values of $u$ and $f$ at grid point $j$, respectively. Finally, the discretized solutions $u_j$ are obtained, by solving the system of linear equations \cref{equation : system of linear equations : periodic boundary condition}.
\begin{gather}
\frac{1}{h^2}
\underbrace
{
\setlength{\arraycolsep}{1.5pt}
\begin{pNiceMatrix}[r,columns-width=auto]
\Block[fill=gray!40,rounded-corners]{2-2}{}
 2 & -1 &     &        &    &    &    &  
\Block[fill=gray!40,rounded-corners]{1-1}{}
                                          -1 \\
-1 &  
\Block[fill=gray!40,rounded-corners]{2-2}{}
      2 &  -1 &        &    &    &    &    \\
   & -1 &   2 &        &    &    &    &    \\
   &    &     & \Ddots &    &    &    &    \\
   &    &     &        &
\Block[fill=gray!40,rounded-corners]{2-2}{} 
                          2 & -1 &    &    \\
   &    &     &        & -1 &
\Block[fill=gray!40,rounded-corners]{2-2}{}
                               2 & -1 &    \\
   &    &     &        &    & -1 &  
\Block[fill=gray!40,rounded-corners]{2-2}{}
                                    2 & -1 \\
\Block[fill=gray!40,rounded-corners]{1-1}{}
-1 &    &      &       &    &    & -1 &  2 
\end{pNiceMatrix}
}
_
{\mathbf{L}_p}
\cdot
\underbrace
{
\begin{pNiceMatrix}[l]
u_0 \\
u_1 \\
u_2 \\
u_3 \\
\Vdots \\
u_{N-3} \\
u_{N-2} \\
u_{N-1}
\end{pNiceMatrix}
}
_
{\ket{u}_p}
=
\underbrace
{
\begin{pNiceMatrix}[l]
f_0 \\
f_1 \\
f_2 \\
f_3 \\
\Vdots \\
f_{N-3} \\
f_{N-2} \\
f_{N-1}
\end{pNiceMatrix}
}
_
{\ket{f}_p}
\label{equation : system of linear equations : periodic boundary condition}
\end{gather}

\subsection{Boundary Conditions}
\label{subsection : preliminaries : boundary conditions}

Similarly, we can form different boundary-value problems by replacing the periodic boundary condition in \cref{equation : periodic boundary condition} by other boundary conditions including Dirichlet, Neumann, and Robin as given in \cref{equation : Dirichlet boundary condition,equation : Neumann boundary condition,equation : Robin boundary condition} respectively, with $a,b,c,d,A,B \in \mathbb{R}$ such that the problems are stable, consistent, and thus converged to the actual solution for large $N$. Their corresponding systems of linear equations are given in \cref{equation : system of linear equations : Dirichlet boundary condition,equation : system of linear equations : Neumann boundary condition,equation : system of linear equations : Robin boundary condition} using the same discretization grid and approximation scheme, except in the Dirichlet case, which use $N+2$ gird points and $h := 1/(N+2)$ grid size instead.
\begin{align}
u(0) = a
\quad
&\text{and}
\quad
u(1) = b
\label{equation : Dirichlet boundary condition}
\\
u'(0) = a
\quad
&\text{and}
\quad
u'(1) = b
\label{equation : Neumann boundary condition}
\\
au(0) + bu'(0) = A
\quad
&\text{and}
\quad
cu(1) + du'(1) = B
\label{equation : Robin boundary condition}
\end{align}
\begin{gather}
\frac{1}{h^2}
\underbrace
{
\setlength{\arraycolsep}{1.5pt}
\begin{pNiceMatrix}[r,columns-width=auto]
\Block[fill=gray!40,rounded-corners]{2-2}{}
 2 & -1 &     &        &    &    &    &  
\Block[fill=gray!40,rounded-corners]{1-1}{}
                                          0 \\
-1 &  
\Block[fill=gray!40,rounded-corners]{2-2}{}
      2 &  -1 &        &    &    &    &    \\
   & -1 &   2 &        &    &    &    &    \\
   &    &     & \Ddots &    &    &    &    \\
   &    &     &        &   
\Block[fill=gray!40,rounded-corners]{2-2}{}
                          2 & -1 &    &    \\
   &    &     &        & -1 &
\Block[fill=gray!40,rounded-corners]{2-2}{}
                               2 & -1 &    \\
   &    &     &        &    & -1 &  
\Block[fill=gray!40,rounded-corners]{2-2}{}
                                    2 & -1 \\
\Block[fill=gray!40,rounded-corners]{1-1}{}
 0 &    &      &       &    &    & -1 &  2 
\end{pNiceMatrix}
}
_
{\mathbf{L}_D}
\cdot
\underbrace
{
\begin{pNiceMatrix}[l]
u_1 \\
u_2 \\
u_3 \\
u_4 \\
\Vdots \\
u_{N-2} \\
u_{N-1} \\
u_{N}
\end{pNiceMatrix}
}
_
{\ket{u}_D}
=
\underbrace
{
\begin{pNiceMatrix}[l]
f_1 - a/h^2 \\
f_2 \\
f_3 \\
f_4 \\
\Vdots \\
f_{N-2} \\
f_{N-1} \\
f_N - b/h^2
\end{pNiceMatrix}
}
_
{\ket{f}_D}
\label{equation : system of linear equations : Dirichlet boundary condition}
\\[.5cm]
\frac{1}{h^2}
\underbrace
{
\setlength{\arraycolsep}{1.5pt}
\begin{pNiceMatrix}[r,columns-width=auto]
\Block[fill=gray!40,rounded-corners]{2-2}{}
 1 & -1 &     &        &    &    &    &  
\Block[fill=gray!40,rounded-corners]{1-1}{}
                                          0 \\
-1 &  
\Block[fill=gray!40,rounded-corners]{2-2}{}
      2 &  -1 &        &    &    &    &    \\
   & -1 &   2 &        &    &    &    &    \\
   &    &     & \Ddots &    &    &    &    \\
   &    &     &        &   
\Block[fill=gray!40,rounded-corners]{2-2}{}
                          2 & -1 &    &    \\
   &    &     &        & -1 &
\Block[fill=gray!40,rounded-corners]{2-2}{}
                               2 & -1 &    \\
   &    &     &        &    & -1 &  
\Block[fill=gray!40,rounded-corners]{2-2}{}
                                    2 & -1 \\
\Block[fill=gray!40,rounded-corners]{1-1}{}
 0 &    &      &       &    &    & -1 &  1 
\end{pNiceMatrix}
}
_
{\mathbf{L}_N}
\cdot
\underbrace
{
\begin{pNiceMatrix}[l]
u_0 \\
u_1 \\
u_2 \\
u_3 \\
\Vdots \\
u_{N-3} \\
u_{N-2} \\
u_{N-1}
\end{pNiceMatrix}
}
_
{\ket{u}_N}
=
\underbrace
{
\begin{pNiceMatrix}[l]
-a/h \\
f_1 \\
f_2 \\
f_3 \\
\Vdots \\
f_{N-3} \\
f_{N-2} \\
b/h
\end{pNiceMatrix}
}
_
{\ket{f}_N}
\label{equation : system of linear equations : Neumann boundary condition}
\\[.5cm]
\frac{1}{h^2}
\underbrace
{
\setlength{\arraycolsep}{1.5pt}
\begin{pNiceMatrix}[r,columns-width=auto]
\Block[fill=gray!40,rounded-corners]{2-2}{}
 C & -1 &     &        &    &    &    &  
\Block[fill=gray!40,rounded-corners]{1-1}{}
                                          0 \\
-1 &  
\Block[fill=gray!40,rounded-corners]{2-2}{}
      2 &  -1 &        &    &    &    &    \\
   & -1 &   2 &        &    &    &    &    \\
   &    &     & \Ddots &    &    &    &    \\
   &    &     &        &   
\Block[fill=gray!40,rounded-corners]{2-2}{}
                          2 & -1 &    &    \\
   &    &     &        & -1 &
\Block[fill=gray!40,rounded-corners]{2-2}{}
                               2 & -1 &    \\
   &    &     &        &    & -1 &  
\Block[fill=gray!40,rounded-corners]{2-2}{}
                                    2 & -1 \\
\Block[fill=gray!40,rounded-corners]{1-1}{}
 0 &    &      &       &    &    & -1 &  D 
\end{pNiceMatrix}
}
_
{\mathbf{L}_R}
\cdot
\underbrace
{
\begin{pNiceMatrix}[l]
u_0 \\
u_1 \\
u_2 \\
u_3 \\
\Vdots \\
u_{N-3} \\
u_{N-2} \\
u_{N-1}
\end{pNiceMatrix}
}
_
{\ket{u}_R}
=
\underbrace
{
\begin{pNiceMatrix}[l]
-A/b \\
f_1 \\
f_2 \\
f_3 \\
\Vdots \\
f_{N-3} \\
f_{N-2} \\
B/d
\end{pNiceMatrix}
}
_
{\ket{f}_R}
\label{equation : system of linear equations : Robin boundary condition}
\shortintertext{where}
C := 1 + ah/b
\quad
\text{and}
\quad
D := 1 + ch/d
\label{equation : placeholder : system of linear equations : Robin boundary condition}
\end{gather}
and the subscripts $p$, $D$, $N$, and $R$ stand for periodic, Dirichlet, Neumann, and Robin, respectively.

\subsection{Higher Dimensions}
\label{subsection : preliminaries : higher dimensions}

Additionally, we can also extend the boundary-value problem to an arbitrary dimension $d \in \mathbb{N}$, by replacing \cref{equation : Poisson's equation} with a generalized Poisson's equation in \cref{equation : Poisson's equation : higher dimensions} where $u, f : [0,1]^d \rightarrow \mathbb{R}$, and imposing the same boundary condition in each dimension. Note that we also assume that the boundary condition in each dimension is separable from each other.

Using the same discretization grid and approximation scheme in each dimension, we derive a system of linear equations $\mathcal{L}\ket{u} = \ket{f}$, where $\ket{u}$ and $\ket{f}$ comprises the discretized values of $u$ and $f$ at grid point $j$ for all $\smash{j \in \{0,\ldots,N-1\}^d}$. \Cref{equation : system of linear equations : higher dimensions,equation : placeholder : system of linear equations : higher dimensions} give the definition of $\mathcal{L}$, as a function of a one-dimensional matrix $\mathbf{L} \in \{\mathbf{L}_p,\mathbf{L}_D,\mathbf{L}_N,\mathbf{L}_R\}$.
\begin{gather}
-\sum_{k=1}^d
\frac{\partial^2}{\partial x_k^2}
u(\mathbf{x})
=
f(\mathbf{x})
\quad
\text{for all}
\quad
\mathbf{x} \in (0,1)^d
\label{equation : Poisson's equation : higher dimensions}
\\
\mathcal{L}
:=
\sum_{k=1}^d
\mathcal{L}_k
\label{equation : system of linear equations : higher dimensions}
\shortintertext{where}
\mathcal{L}_k
:=
\mathbf{I}_N^{\otimes (k-1)}
\otimes
\mathbf{L}
\otimes
\mathbf{I}_N^{\otimes (d-k)}
\quad
\text{and}
\quad
\mathbf{L}
\in
\{
\mathbf{L}_p,
\mathbf{L}_D,
\mathbf{L}_N,
\mathbf{L}_R
\}
\label{equation : placeholder : system of linear equations : higher dimensions}
\end{gather}

It is also possible to have a mixed boundary condition, e.g., a Dirichlet boundary condition on one side of the boundary in a particular dimension, and a Neumann boundary condition on the other side of the boundary in the same dimension or a different dimension. However, the domain may become hyper-rectangular instead of the hypercube that we have.

Additionally, note that all the matrices $\mathbf{L}_p$, $\mathbf{L}_D$, $\mathbf{L}_N$, $\mathbf{L}_R$ and $\mathcal{L}$ are all Hermitian and positive definite, which means that they are invertible, and their corresponding systems of linear equations have a unique solution.

\section{Block-Diagonalization}
\label{section : block-diagonalization}

The primary objective of this article is to show that the matrices $\mathbf{L}_p$, $\mathbf{L}_D$, $\mathbf{L}_N$ and $\mathbf{L}_R$ which are derived using finite difference method for some boundary-value problems, can be block-encoded inside a unitary operator with a double-logarithmic depth $\mathcal{O}(\log(\log(N)))$ in their matrix size $N$. To achieve such objective, we introduce the concept of a \emph{block-diagonalizable} matrix in \cref{definition : block-diagonalizable matrix}, and show that all the aforementioned matrices are block-diagonalizable.

\begin{definition}[Block-Diagonalizable Matrix]
\label{definition : block-diagonalizable matrix}
An $N \times N$ matrix $\mathbf{L}$ is \emph{block-diagonalizable} if
\begin{equation}
\mathbf{L}
=
\sum_{c \in \{0,1\}}
\sum_{\hat{\sigma} \in \{\mathbf{I},\mathbf{X},\mathbf{Y},\mathbf{Z}\}}
\chi_{\hat{\sigma}}
\cdot
\mathbf{L}_{c,\hat{\sigma}}
\end{equation}
for some
\begin{align}
\chi_{\hat{\sigma}} 
&\in 
\mathbb{R}
\\
\mathbf{L}_{c,\hat{\sigma}} 
&:= 
\mathbf{P}_c^{\vphantom{\dagger}} 
\cdot 
\mathbf{D}_{\hat{\sigma}} 
\cdot 
\mathbf{P}_c^{\dagger}
\\
\mathbf{D}_{\hat{\sigma}}
&:=
\diag(\alpha_0,\ldots,\alpha_{N/2-1})_{N/2}
\otimes
\hat{\sigma}
\\
\mathbf{P}_c
&:=
\sum_{i=0}^{N-1}
\ketbra{i + c \bmod N}{i}
\end{align}
where $\mathbf{D}_{\hat{\sigma}}$ and $\mathbf{P}_c$ are called Pauli-block-diagonal matrix and permutation matrix, respectively.
\end{definition}

In summary, a block-diagonalizable matrix $\mathbf{L}$ can be decomposed as a linear combination of linear operators $\mathbf{L}_{c,\hat{\sigma}}$, for some $c \in \{0,1\}$ and Pauli matrix $\hat{\sigma}$; each of which is composed of a Pauli-block-diagonal matrix $\mathbf{D}_{\hat{\sigma}}$ that is left and right multiplied by a permutation matrix $\smash{\mathbf{P}_c^{\vphantom{\dagger}}}$ and its conjugate transpose $\smash{\mathbf{P}_c^{\dagger}}$, respectively. Such a matrix decomposition process is called \emph{block-diagonalization}; from which a procedure for constructing a block-encoding matrix is derived.

This methodology is inspired by the combinatorially block-diagonal matrix definition of \cite{aharonov2003adiabatic}, since each of our linear operators $\mathbf{L}_{c,\hat{\sigma}}$ is in fact a $2 \times 2$ combinatorially block-diagonal matrix, where the permutation matrices $\smash{\mathbf{P}_c^{\vphantom{\dagger}}}$ and $\smash{\mathbf{P}_c^\dagger}$ are the combinatorial permutations. Our block-diagonalizable matrix definition is basically a linear combination of $2 \times 2$ combinatorially block-diagonal matrices; each of which is a tensor-product of a diagonal matrix and a Pauli matrix.

This section is divided into five subsections, which describe the block-diagonalization of $\mathbf{L}_p$ with periodic boundary condition in \cref{subsection : block-diagonalization : periodic boundary condition}, $\mathbf{L}_D$ with Dirichlet boundary condition in \cref{subsection : block-diagonalization : Dirichlet boundary condition}, $\mathbf{L}_N$ with Neumann boundary condition in \cref{subsection : block-diagonalization : Neumann boundary condition}, $\mathbf{L}_R$ with Robin boundary condition in \cref{subsection : block-diagonalization : Robin boundary condition}, and $\mathcal{L}$ from higher dimensional problems in \cref{subsection : block-diagonalization : higher dimensions}.  

\subsection{Periodic}
\label{subsection : block-diagonalization : periodic boundary condition}

We begin by decomposing $\mathbf{L}_p$ as a sum of two matrices $\mathbf{L}_{0}$ and $\mathbf{L}_{1}$ as shown in \cref{equation : system decomposition : periodic boundary condition}.
\begin{gather}
\underbrace
{
\setlength{\arraycolsep}{1.5pt}
\begin{pNiceMatrix}[r,columns-width=auto]
\Block[fill=gray!40,rounded-corners]{2-2}{}
 2 & -1 &     &        &    &    &    &  
\Block[fill=gray!40,rounded-corners]{1-1}{}
                                        -1 \\
-1 &  
\Block[fill=gray!40,rounded-corners]{2-2}{}
      2 &  -1 &        &    &    &    &    \\
   & -1 &   2 &        &    &    &    &    \\
   &    &     & \Ddots &    &    &    &    \\
   &    &     &        &
\Block[fill=gray!40,rounded-corners]{2-2}{} 
                          2 & -1 &    &    \\
   &    &     &        & -1 &
\Block[fill=gray!40,rounded-corners]{2-2}{}
                               2 & -1 &    \\
   &    &     &        &    & -1 &  
\Block[fill=gray!40,rounded-corners]{2-2}{}
                                    2 & -1 \\
\Block[fill=gray!40,rounded-corners]{1-1}{}
-1 &    &      &       &    &    & -1 &  2 
\end{pNiceMatrix}
}
_
{\mathbf{L}_p}
=
\underbrace
{
\setlength{\arraycolsep}{1.5pt}
\begin{pNiceMatrix}[r,columns-width=auto]
\Block[fill=gray!40,rounded-corners]{2-2}{}
 1 & -1 &        &    &    &    &    &     \\
-1 &  1 &        &    &    &    &    &     \\
   &    & \Ddots &    &    &    &    &    \\
   &    &        &    &    &    &    &    \\
   &    &        &    &  
\Block[fill=gray!40,rounded-corners]{2-2}{}
                         1 & -1 &    &    \\
   &    &        &    & -1 &  1 &    &    \\
   &    &        &    &    &    &  
\Block[fill=gray!40,rounded-corners]{2-2}{}
                                   1 & -1 \\
   &    &      &    &      &    & -1 &  1 
\end{pNiceMatrix}
}
_
{\mathbf{L}_{0}}
+
\underbrace
{
\setlength{\arraycolsep}{1.5pt}
\begin{pNiceMatrix}[r,columns-width=auto]
\Block[fill=gray!40,rounded-corners]{1-1}{}
 1 &    &    &        &   &    &    &
\Block[fill=gray!40,rounded-corners]{1-1}{}
                                      -1 \\
   &  
\Block[fill=gray!40,rounded-corners]{2-2}{}
      1 & -1 &        &   &    &    &    \\
   & -1 &  1 &        &   &    &    &    \\
   &    &    & \Ddots &   &    &    &    \\
   &    &    &        &   &    &    &    \\
   &    &    &        &   & 
\Block[fill=gray!40,rounded-corners]{2-2}{}   
                             1 & -1 &    \\
   &    &    &        &   & -1 &  1 &    \\
\Block[fill=gray!40,rounded-corners]{1-1}{}
-1 &    &    &        &   &    &    & 
\Block[fill=gray!40,rounded-corners]{1-1}{}
                                       1 
\end{pNiceMatrix}
}
_
{\mathbf{L}_{1}}
\label{equation : system decomposition : periodic boundary condition}
\end{gather}

We define a $2 \times 2$ block-diagonal matrix $\mathbf{B}$ as shown in \cref{equation : block-diagonal matrices : periodic boundary condition}. Then, we show that it can be transformed into $\mathbf{L}_{0}$ and $\mathbf{L}_{1}$ as in \cref{equation : subsystem decomposition : periodic boundary condition}, where $\mathbf{P}_0$ and $\mathbf{P}_1$ are permutation matrices as defined in \cref{equation : permutation matrices : periodic boundary condition}. As in \cite{aharonov2003adiabatic}, we say that $\mathbf{L}_{0}$ and $\mathbf{L}_{1}$ are $2 \times 2$ combinatorially block-diagonal matrices, i.e., block-diagonal given some row and column rearrangements.
\begin{gather}
\mathbf{B}
:=
{
\setlength{\arraycolsep}{1.5pt}
\begin{pNiceMatrix}[r,columns-width=auto]
\Block[fill=gray!40,rounded-corners]{2-2}{}
 1 & -1 &        &    &    &    &    &     \\
-1 &  1 &        &    &    &    &    &     \\
   &    & \Ddots &    &    &    &    &    \\
   &    &        &    &    &    &    &    \\
   &    &        &    &  
\Block[fill=gray!40,rounded-corners]{2-2}{}
                         1 & -1 &    &    \\
   &    &        &    & -1 &  1 &    &    \\
   &    &        &    &    &    &  
\Block[fill=gray!40,rounded-corners]{2-2}{}
                                   1 & -1 \\
   &    &      &    &      &    & -1 &  1 
\end{pNiceMatrix}
}
\label{equation : block-diagonal matrices : periodic boundary condition}
\\[.3cm]
\mathbf{L}_{0}
=
\mathbf{P}_0
\cdot
\mathbf{B}
\cdot
\mathbf{P}_0^\dagger
\quad
\text{and}
\quad
\mathbf{L}_{1}
=
\mathbf{P}_1
\cdot
\mathbf{B}
\cdot
\mathbf{P}_1^\dagger
\label{equation : subsystem decomposition : periodic boundary condition}
\\[.3cm]
\mathbf{P}_0
:=
\sum_{i=0}^{N-1}
\ketbra{i}{i}
\quad
\text{and}
\quad
\mathbf{P}_1
:=
\sum_{i=0}^{N-1}
\ketbra{i + 1 \bmod N}{i}
\label{equation : permutation matrices : periodic boundary condition}
\end{gather}

We decompose the $2 \times 2$ block-matrix $((1,-1),(-1,1))$ as a sum of Pauli matrices $\mathbf{I}$ and $\mathbf{X}$ as in \cref{equation : block matrix decomposition : periodic boundary condition}. Substituting that into \cref{equation : block-diagonal matrices : periodic boundary condition}, we can rewrite $\mathbf{B}$ as the sum of two Pauli-block-diagonal matrices $\mathbf{D}_{\mathbf{I}}$ and $\mathbf{D}_{\mathbf{X}}$ as in \cref{equation : block-diagonal matrix decomposition : periodic boundary condition}.
\begin{gather}
{
\setlength{\arraycolsep}{1.5pt}
\begin{pNiceMatrix}[r,columns-width=auto]
 1 & -1 \\
-1 &  1 
\end{pNiceMatrix}
}
=
\underbrace
{
\begin{pNiceMatrix}[r]
1 & 0 \\
0 & 1 
\end{pNiceMatrix}
}
_
{\mathbf{I}}
-
\underbrace
{
\begin{pNiceMatrix}[r]
0 & 1 \\
1 & 0 
\end{pNiceMatrix}
}
_
{\mathbf{X}}
\label{equation : block matrix decomposition : periodic boundary condition}
\\[.3cm]
\mathbf{B}
=
\underbrace
{
\diag(1,\ldots,1)_{N/2}
\otimes
\mathbf{I}
}
_
{\mathbf{D}_{\mathbf{I}}}
-
\underbrace
{
\diag(1,\ldots,1)_{N/2}
\otimes
\mathbf{X}
}
_
{\mathbf{D}_{\mathbf{X}}}
\label{equation : block-diagonal matrix decomposition : periodic boundary condition}
\end{gather}

Finally, substituting \cref{equation : block-diagonal matrix decomposition : periodic boundary condition} into \cref{equation : subsystem decomposition : periodic boundary condition}, and again into \cref{equation : system decomposition : periodic boundary condition}, we obtain a block-diagonalization of $\mathbf{L}_p$ in \cref{theorem : block-diagonalization : periodic boundary condition}.

\begin{restatable}[Block-Diagonalization, Periodic]{theorem}{theorem : block-diagonalization : periodic boundary condition}
\label{theorem : block-diagonalization : periodic boundary condition}
\begin{equation}
\mathbf{L}_p
=
\sum_{c \in \{0,1\}}
\sum_{\hat{\sigma} \in \{\mathbf{I},\mathbf{X}\}}
\chi_{\hat{\sigma}}
\cdot
\mathbf{L}_{c, \hat{\sigma}}
\label{equation : block-diagonalization : periodic boundary condition}
\end{equation}
where
\begin{align}
\chi_{\hat{\sigma}}
&
:=
\begin{cases}
\hphantom{-}1, & \hat{\sigma} = \mathbf{I} \\
-1, & \hat{\sigma} = \mathbf{X}
\end{cases}
\label{equation : coefficient : periodic boundary condition}
\\
\mathbf{L}_{c, \hat{\sigma}}
&
:=
\mathbf{P}_c
\cdot
\mathbf{D}_{\hat{\sigma}}
\cdot
\mathbf{P}_c^\dagger
\label{equation : subsystem : periodic boundary condition}
\\
\mathbf{D}_{\hat{\sigma}}
&
:=
\diag(1, \ldots, 1)_{N/2}
\otimes
\hat{\sigma}
\label{equation : block-diagonal : periodic boundary condition}
\\
\mathbf{P}_c
&
:=
\sum_{i=0}^{N-1}
\ketbra{i + c \bmod N}{i}
\label{equation : permutation : periodic boundary condition}
\end{align}
\end{restatable}

\subsection{Dirichlet}
\label{subsection : block-diagonalization : Dirichlet boundary condition}

The following derives block-diagonalization of $\mathbf{L}_D$ as in \cref{theorem : block-diagonalization : Dirichlet boundary condition}, using similar procedures as those of the periodic case. However, instead of defining a single $2 \times 2$ block-diagonal matrix $\mathbf{B}$, we use two such matrices $\mathbf{B}_{0}$ and $\mathbf{B}_{1}$ instead. Additionally, one of the $2 \times 2$ block matrices $((1,0),(0,1))$ is different, and also has a different Pauli decomposition.
\begin{gather}
\underbrace
{
\setlength{\arraycolsep}{1.5pt}
\begin{pNiceMatrix}[r,columns-width=auto]
\Block[fill=gray!40,rounded-corners]{2-2}{}
 2 & -1 &     &        &    &    &    &  
\Block[fill=gray!40,rounded-corners]{1-1}{}
                                         0 \\
-1 &  
\Block[fill=gray!40,rounded-corners]{2-2}{}
      2 &  -1 &        &    &    &    &    \\
   & -1 &   2 &        &    &    &    &    \\
   &    &     & \Ddots &    &    &    &    \\
   &    &     &        &   
\Block[fill=gray!40,rounded-corners]{2-2}{}
                          2 & -1 &    &    \\
   &    &     &        & -1 &
\Block[fill=gray!40,rounded-corners]{2-2}{}
                               2 & -1 &    \\
   &    &     &        &    & -1 &  
\Block[fill=gray!40,rounded-corners]{2-2}{}
                                    2 & -1 \\
\Block[fill=gray!40,rounded-corners]{1-1}{}
 0 &    &      &       &    &    & -1 &  2 
\end{pNiceMatrix}
}
_
{\mathbf{L}_D}
=
\underbrace
{
\setlength{\arraycolsep}{1.5pt}
\begin{pNiceMatrix}[r,columns-width=auto]
\Block[fill=gray!40,rounded-corners]{2-2}{}
 1 & -1 &        &    &    &    &    &     \\
-1 &  1 &        &    &    &    &    &     \\
   &    & \Ddots &    &    &    &    &    \\
   &    &        &    &    &    &    &    \\
   &    &        &    &  
\Block[fill=gray!40,rounded-corners]{2-2}{}
                         1 & -1 &    &    \\
   &    &        &    & -1 &  1 &    &    \\
   &    &        &    &    &    &  
\Block[fill=gray!40,rounded-corners]{2-2}{}
                                   1 & -1 \\
   &    &      &    &      &    & -1 &  1 
\end{pNiceMatrix}
}
_
{\mathbf{L}_{0}}
+
\underbrace
{
\setlength{\arraycolsep}{1.5pt}
\begin{pNiceMatrix}[r,columns-width=auto]
\Block[fill=gray!40,rounded-corners]{1-1}{}
 1 &    &    &        &   &    &    &
\Block[fill=gray!40,rounded-corners]{1-1}{}
                                       0 \\
   &  
\Block[fill=gray!40,rounded-corners]{2-2}{}
      1 & -1 &        &   &    &    &    \\
   & -1 &  1 &        &   &    &    &    \\
   &    &    & \Ddots &   &    &    &    \\
   &    &    &        &   &    &    &    \\
   &    &    &        &   & 
\Block[fill=gray!40,rounded-corners]{2-2}{}   
                             1 & -1 &    \\
   &    &    &        &   & -1 &  1 &    \\
\Block[fill=gray!40,rounded-corners]{1-1}{}
 0 &    &    &        &   &    &    & 
\Block[fill=gray!40,rounded-corners]{1-1}{}
                                       1 
\end{pNiceMatrix}
}
_
{\mathbf{L}_{1}}
\label{equation : system decomposition : Dirichlet boundary condition}
\\[.5cm]
\mathbf{B}_{0}
:=
{
\setlength{\arraycolsep}{1.5pt}
\begin{pNiceMatrix}[r,columns-width=auto]
\Block[fill=gray!40,rounded-corners]{2-2}{}
 1 & -1 &        &    &    &    &    &     \\
-1 &  1 &        &    &    &    &    &     \\
   &    & \Ddots &    &    &    &    &    \\
   &    &        &    &    &    &    &    \\
   &    &        &    &  
\Block[fill=gray!40,rounded-corners]{2-2}{}
                         1 & -1 &    &    \\
   &    &        &    & -1 &  1 &    &    \\
   &    &        &    &    &    &  
\Block[fill=gray!40,rounded-corners]{2-2}{}
                                   1 & -1 \\
   &    &      &    &      &    & -1 &  1 
\end{pNiceMatrix}
}
\quad
\mathbf{B}_{1}
:=
{
\setlength{\arraycolsep}{1.5pt}
\begin{pNiceMatrix}[r,columns-width=auto]
\Block[fill=gray!40,rounded-corners]{2-2}{}
 1 & -1 &        &    &    &    &    &     \\
-1 &  1 &        &    &    &    &    &     \\
   &    & \Ddots &    &    &    &    &    \\
   &    &        &    &    &    &    &    \\
   &    &        &    &  
\Block[fill=gray!40,rounded-corners]{2-2}{}
                         1 & -1 &    &    \\
   &    &        &    & -1 &  1 &    &    \\
   &    &        &    &    &    &  
\Block[fill=gray!40,rounded-corners]{2-2}{}
                                   1 &  0 \\
   &    &      &    &      &    &  0 &  1 
\end{pNiceMatrix}
}
\label{equation : block-diagonal matrices : Dirichlet boundary condition}
\\[.3cm]
\mathbf{L}_{0}
=
\mathbf{P}_0
\cdot
\mathbf{B}_{0}
\cdot
\mathbf{P}_0^\dagger
\quad
\text{and}
\quad
\mathbf{L}_{1}
=
\mathbf{P}_1
\cdot
\mathbf{B}_{1}
\cdot
\mathbf{P}_1^\dagger
\label{equation : subsystem decomposition : Dirichlet boundary condition}
\\[.3cm]
\begin{aligned}
\mathbf{B}_{0}
&
=
\underbrace
{
\diag(1,\ldots,1,1)_{N/2}
\otimes
\mathbf{I}
}
_
{\mathbf{D}_{0,\mathbf{I}}}
-
\underbrace
{
\diag(1,\ldots,1,1)_{N/2}
\otimes
\mathbf{X}
}
_
{\mathbf{D}_{0,\mathbf{X}}}
\\
\mathbf{B}_{1}
&
=
\underbrace
{
\diag(1,\ldots,1,1)_{N/2}
\otimes
\mathbf{I}
}
_
{\mathbf{D}_{1,\mathbf{I}}}
-
\underbrace
{
\diag(1,\ldots,1,0)_{N/2}
\otimes
\mathbf{X}
}
_
{\mathbf{D}_{1,\mathbf{X}}}
\end{aligned}
\label{equation : block-diagonal matrix decomposition : Dirchlet boundary condition}
\end{gather}

\begin{restatable}[Block-Diagonalization, Dirichlet]{theorem}{theorem : block-diagonalization : Dirichlet boundary condition}
\label{theorem : block-diagonalization : Dirichlet boundary condition}
\begin{equation}
\mathbf{L}_D
=
\sum_{c \in \{0,1\}}
\sum_{\hat{\sigma} \in \{\mathbf{I},\mathbf{X}\}}
\chi_{\hat{\sigma}}
\cdot
\mathbf{L}_{c, \hat{\sigma}}
\label{equation : block-diagonalization : Dirichlet boundary condition}
\end{equation}
where
\begin{align}
\mathbf{L}_{c, \hat{\sigma}}
&
:=
\mathbf{P}_c
\cdot
\mathbf{D}_{c,\hat{\sigma}}
\cdot
\mathbf{P}_c^\dagger
\label{equation : subsystem : Dirichlet boundary condition}
\\
\mathbf{D}_{c,\hat{\sigma}}
&
:=
\begin{cases}
\diag(1, \ldots, 1, 1)_{N/2}
\otimes
\hat{\sigma},
&
(c,\hat{\sigma})
\neq
(1,\mathbf{X})
\\
\diag(1, \ldots, 1, 0)_{N/2}
\otimes
\hat{\sigma},
&
(c,\hat{\sigma})
=
(1,\mathbf{X})
\end{cases}
\label{equation : block-diagonal : Dirichlet boundary condition}
\end{align}
\end{restatable}

\subsection{Neumann}
\label{subsection : block-diagonalization : Neumann boundary condition}

The following derives block-diagonalization of $\mathbf{L}_N$ as in \cref{theorem : block-diagonalization : Neumann boundary condition}, using similar procedures as those of the periodic case. However, instead of defining a single $2 \times 2$ block-diagonal matrix $\mathbf{B}$, we use two such matrices $\mathbf{B}_{0}$ and $\mathbf{B}_{1}$ instead. Additionally, one of the $2 \times 2$ block matrix $((0,0),(0,0))$ is different, and also has a different Pauli decomposition.
\begin{gather}
\underbrace
{
\setlength{\arraycolsep}{1.5pt}
\begin{pNiceMatrix}[r,columns-width=auto]
\Block[fill=gray!40,rounded-corners]{2-2}{}
 1 & -1 &     &        &    &    &    &  
\Block[fill=gray!40,rounded-corners]{1-1}{}
                                          0 \\
-1 &  
\Block[fill=gray!40,rounded-corners]{2-2}{}
      2 &  -1 &        &    &    &    &    \\
   & -1 &   2 &        &    &    &    &    \\
   &    &     & \Ddots &    &    &    &    \\
   &    &     &        &   
\Block[fill=gray!40,rounded-corners]{2-2}{}
                          2 & -1 &    &    \\
   &    &     &        & -1 &
\Block[fill=gray!40,rounded-corners]{2-2}{}
                               2 & -1 &    \\
   &    &     &        &    & -1 &  
\Block[fill=gray!40,rounded-corners]{2-2}{}
                                    2 & -1 \\
\Block[fill=gray!40,rounded-corners]{1-1}{}
 0 &    &      &       &    &    & -1 &  1 
\end{pNiceMatrix}
}
_
{\mathbf{L}_N}
=
\underbrace
{
\setlength{\arraycolsep}{1.5pt}
\begin{pNiceMatrix}[r,columns-width=auto]
\Block[fill=gray!40,rounded-corners]{2-2}{}
 1 & -1 &        &    &    &    &    &     \\
-1 &  1 &        &    &    &    &    &     \\
   &    & \Ddots &    &    &    &    &    \\
   &    &        &    &    &    &    &    \\
   &    &        &    &  
\Block[fill=gray!40,rounded-corners]{2-2}{}
                         1 & -1 &    &    \\
   &    &        &    & -1 &  1 &    &    \\
   &    &        &    &    &    &  
\Block[fill=gray!40,rounded-corners]{2-2}{}
                                   1 & -1 \\
   &    &      &    &      &    & -1 &  1 
\end{pNiceMatrix}
}
_
{\mathbf{L}_{0}}
+
\underbrace
{
\setlength{\arraycolsep}{1.5pt}
\begin{pNiceMatrix}[r,columns-width=auto]
\Block[fill=gray!40,rounded-corners]{1-1}{}
 0 &    &    &        &   &    &    &
\Block[fill=gray!40,rounded-corners]{1-1}{}
                                       0 \\
   &  
\Block[fill=gray!40,rounded-corners]{2-2}{}
      1 & -1 &        &   &    &    &    \\
   & -1 &  1 &        &   &    &    &    \\
   &    &    & \Ddots &   &    &    &    \\
   &    &    &        &   &    &    &    \\
   &    &    &        &   & 
\Block[fill=gray!40,rounded-corners]{2-2}{}   
                             1 & -1 &    \\
   &    &    &        &   & -1 &  1 &    \\
\Block[fill=gray!40,rounded-corners]{1-1}{}
 0 &    &    &        &   &    &    & 
\Block[fill=gray!40,rounded-corners]{1-1}{}
                                       0 
\end{pNiceMatrix}
}
_
{\mathbf{L}_{1}}
\label{equation : system decomposition : Neumann boundary condition}
\\[.5cm]
\mathbf{B}_{0}
:=
{
\setlength{\arraycolsep}{1.5pt}
\begin{pNiceMatrix}[r,columns-width=auto]
\Block[fill=gray!40,rounded-corners]{2-2}{}
 1 & -1 &        &    &    &    &    &     \\
-1 &  1 &        &    &    &    &    &     \\
   &    & \Ddots &    &    &    &    &    \\
   &    &        &    &    &    &    &    \\
   &    &        &    &  
\Block[fill=gray!40,rounded-corners]{2-2}{}
                         1 & -1 &    &    \\
   &    &        &    & -1 &  1 &    &    \\
   &    &        &    &    &    &  
\Block[fill=gray!40,rounded-corners]{2-2}{}
                                   1 & -1 \\
   &    &      &    &      &    & -1 &  1 
\end{pNiceMatrix}
}
\quad
\mathbf{B}_{1}
:=
{
\setlength{\arraycolsep}{1.5pt}
\begin{pNiceMatrix}[r,columns-width=auto]
\Block[fill=gray!40,rounded-corners]{2-2}{}
 1 & -1 &        &    &    &    &    &     \\
-1 &  1 &        &    &    &    &    &     \\
   &    & \Ddots &    &    &    &    &    \\
   &    &        &    &    &    &    &    \\
   &    &        &    &  
\Block[fill=gray!40,rounded-corners]{2-2}{}
                         1 & -1 &    &    \\
   &    &        &    & -1 &  1 &    &    \\
   &    &        &    &    &    &  
\Block[fill=gray!40,rounded-corners]{2-2}{}
                                   0 &  0 \\
   &    &      &    &      &    &  0 &  0 
\end{pNiceMatrix}
}
\label{equation : block-diagonal matrices : Neumann boundary condition}
\\[.3cm]
\mathbf{L}_{0}
=
\mathbf{P}_0
\cdot
\mathbf{B}_{0}
\cdot
\mathbf{P}_0^\dagger
\quad
\text{and}
\quad
\mathbf{L}_{1}
=
\mathbf{P}_1
\cdot
\mathbf{B}_{1}
\cdot
\mathbf{P}_1^\dagger
\label{equation : subsystem decomposition : Neumann boundary condition}
\\[.3cm]
\begin{aligned}
\mathbf{B}_{0}
&
=
\underbrace
{
\diag(1,\ldots,1,1)_{N/2}
\otimes
\mathbf{I}
}
_
{\mathbf{D}_{0,\mathbf{I}}}
-
\underbrace
{
\diag(1,\ldots,1,1)_{N/2}
\otimes
\mathbf{X}
}
_
{\mathbf{D}_{0,\mathbf{X}}}
\\
\mathbf{B}_{1}
&
=
\underbrace
{
\diag(1,\ldots,1,0)_{N/2}
\otimes
\mathbf{I}
}
_
{\mathbf{D}_{1,\mathbf{I}}}
-
\underbrace
{
\diag(1,\ldots,1,0)_{N/2}
\otimes
\mathbf{X}
}
_
{\mathbf{D}_{1,\mathbf{X}}}
\end{aligned}
\label{equation : block-diagonal matrix decomposition : Neumann boundary condition}
\end{gather}

\begin{restatable}[Block-Diagonalization, Neumann]{theorem}{theorem : block-diagonalization : Neumann boundary condition}
\label{theorem : block-diagonalization : Neumann boundary condition}
\begin{equation}
\mathbf{L}_D
=
\sum_{c \in \{0,1\}}
\sum_{\hat{\sigma} \in \{\mathbf{I},\mathbf{X}\}}
\chi_{\hat{\sigma}}
\cdot
\mathbf{L}_{c, \hat{\sigma}}
\label{equation : block-diagonalization : Neumann boundary condition}
\end{equation}
where
\begin{align}
\mathbf{L}_{c, \hat{\sigma}}
&
:=
\mathbf{P}_c
\cdot
\mathbf{D}_{c,\hat{\sigma}}
\cdot
\mathbf{P}_c^\dagger
\label{equation : subsystem : Neumann boundary condition}
\\
\mathbf{D}_{c,\hat{\sigma}}
&
:=
\begin{cases}
\diag(1, \ldots, 1, 1)_{N/2}
\otimes
\hat{\sigma},
&
c = 0
\\
\diag(1, \ldots, 1, 0)_{N/2}
\otimes
\hat{\sigma},
&
c = 1
\end{cases}
\label{equation : block-diagonal : Neumann boundary condition}
\end{align}
\end{restatable}

\subsection{Robin}
\label{subsection : block-diagonalization : Robin boundary condition}

The following derives block-diagonalization of $\mathbf{L}_R$ as in \cref{theorem : block-diagonalization : Robin boundary condition}, using similar procedures as those of the periodic case. However, instead of defining a single $2 \times 2$ block-diagonal matrix $\mathbf{B}$, we use two such matrices $\mathbf{B}_{0}$ and $\mathbf{B}_{1}$ instead. Additionally, two of the $2 \times 2$ block matrices are different, and thus have different Pauli decompositions, as in \cref{equation : block matrix decomposition : Robin boundary condition}. 
\begin{gather}
\underbrace
{
\setlength{\arraycolsep}{1.5pt}
\begin{pNiceMatrix}[r,columns-width=auto]
\Block[fill=gray!40,rounded-corners]{2-2}{}
 C & -1 &     &        &    &    &    &  
\Block[fill=gray!40,rounded-corners]{1-1}{}
                                          0 \\
-1 &  
\Block[fill=gray!40,rounded-corners]{2-2}{}
      2 &  -1 &        &    &    &    &    \\
   & -1 &   2 &        &    &    &    &    \\
   &    &     & \Ddots &    &    &    &    \\
   &    &     &        &   
\Block[fill=gray!40,rounded-corners]{2-2}{}
                          2 & -1 &    &    \\
   &    &     &        & -1 &
\Block[fill=gray!40,rounded-corners]{2-2}{}
                               2 & -1 &    \\
   &    &     &        &    & -1 &  
\Block[fill=gray!40,rounded-corners]{2-2}{}
                                    2 & -1 \\
\Block[fill=gray!40,rounded-corners]{1-1}{}
 0 &    &      &       &    &    & -1 &  D 
\end{pNiceMatrix}
}
_
{\mathbf{L}_R}
=
\underbrace
{
\setlength{\arraycolsep}{1.5pt}
\begin{pNiceMatrix}[r,columns-width=auto]
\Block[fill=gray!40,rounded-corners]{2-2}{}
 1 & -1 &        &   &    &    &    &    \\
-1 &  1 &        &   &    &    &    &    \\
   &    & \Ddots &   &    &    &    &    \\
   &    &        &   &    &    &    &    \\
   &    &        &   &  
\Block[fill=gray!40,rounded-corners]{2-2}{}
                        1 & -1 &    &    \\
   &    &        &   & -1 &  1 &    &    \\
   &    &        &   &    &    &
\Block[fill=gray!40,rounded-corners]{2-2}{}
                                  2 & -1 \\
   &    &        &   &    &    & -1 & D' 
\end{pNiceMatrix}
}
_
{\mathbf{L}_{0}}
+
\underbrace
{
\setlength{\arraycolsep}{1.5pt}
\begin{pNiceMatrix}[r,columns-width=auto]
\Block[fill=gray!40,rounded-corners]{1-1}{}
C'    &    &    &        &   &    &    &
\Block[fill=gray!40,rounded-corners]{1-1}{}
                                          0 \\
      &  
\Block[fill=gray!40,rounded-corners]{2-2}{}
         1 & -1 &        &   &    &    &    \\
      & -1 &  1 &        &   &    &    &    \\
      &    &    & \Ddots &   &    &    &    \\
      &    &    &        &   &    &    &    \\
      &    &    &        &   & 
\Block[fill=gray!40,rounded-corners]{2-2}{}   
                                1 & -1 &    \\
      &    &    &        &   & -1 &  1 &    \\
\Block[fill=gray!40,rounded-corners]{1-1}{}
    0 &    &    &        &   &    &    & 
\Block[fill=gray!40,rounded-corners]{1-1}{}
                                          1
\end{pNiceMatrix}
}
_
{\mathbf{L}_{1}}
\label{equation : system decomposition : Robin boundary condition}
\shortintertext{where}
C' := C - 1
\quad
\text{and}
\quad
D' := D - 1
\label{equation : placeholder : system decomposition : Robin boundary condition}
\\[.5cm]
\mathbf{B}_{0}
:=
{
\setlength{\arraycolsep}{1.5pt}
\begin{pNiceMatrix}[r,columns-width=auto]
\Block[fill=gray!40,rounded-corners]{2-2}{}
 1 & -1 &        &    &    &    &    &     \\
-1 &  1 &        &    &    &    &    &     \\
   &    & \Ddots &    &    &    &    &    \\
   &    &        &    &    &    &    &    \\
   &    &        &    &  
\Block[fill=gray!40,rounded-corners]{2-2}{}
                         1 & -1 &    &    \\
   &    &        &    & -1 &  1 &    &    \\
   &    &        &    &    &    &  
\Block[fill=gray!40,rounded-corners]{2-2}{}
                                       1 & -1 \\
   &    &      &    &          &    & -1 & D' 
\end{pNiceMatrix}
}
\quad
\mathbf{B}_{1}
:=
{
\setlength{\arraycolsep}{1.5pt}
\begin{pNiceMatrix}[r,columns-width=auto]
\Block[fill=gray!40,rounded-corners]{2-2}{}
 1 & -1 &        &    &    &    &    &     \\
-1 &  1 &        &    &    &    &    &     \\
   &    & \Ddots &    &    &    &    &    \\
   &    &        &    &    &    &    &    \\
   &    &        &    &  
\Block[fill=gray!40,rounded-corners]{2-2}{}
                         1 & -1 &    &    \\
   &    &        &    & -1 &  1 &    &    \\
   &    &        &    &    &    &  
\Block[fill=gray!40,rounded-corners]{2-2}{}
                                  1\hphantom{'} & 0\hphantom{'} \\
   &    &      &    &      &    & 0\hphantom{'} & C' 
\end{pNiceMatrix}
}
\label{equation : block-diagonal matrices : Robin boundary condition}
\\[.3cm]
\mathbf{L}_{0}
=
\mathbf{P}_0
\cdot
\mathbf{B}_{0}
\cdot
\mathbf{P}_0^\dagger
\quad
\text{and}
\quad
\mathbf{L}_{1}
=
\mathbf{P}_1
\cdot
\mathbf{B}_{1}
\cdot
\mathbf{P}_1^\dagger
\label{equation : subsystem decomposition : Robin boundary condition}
\\[.5cm]
\begin{aligned}
{
\setlength{\arraycolsep}{1.5pt}
\begin{pNiceMatrix}[r,columns-width=auto]
\hphantom{-}1 & -1 \\
-1 & D' 
\end{pNiceMatrix}
}
&
=
\frac{D}{2}
\underbrace
{
\begin{pNiceMatrix}[r]
1 & 0 \\
0 & 1 
\end{pNiceMatrix}
}
_
{\mathbf{I}}
+
\left(
1 - \frac{D}{2}
\right)
\underbrace
{
\setlength{\arraycolsep}{2pt}
\begin{pNiceMatrix}[r]
1 &  0 \\
0 & -1 
\end{pNiceMatrix}
}
_
{\mathbf{Z}}
-
\underbrace
{
\setlength{\arraycolsep}{2pt}
\begin{pNiceMatrix}[r]
0 & \hphantom{-}1 \\
1 & \hphantom{-}0 
\end{pNiceMatrix}
}
_
{\mathbf{X}}
\\
{
\setlength{\arraycolsep}{1.5pt}
\begin{pNiceMatrix}[r,columns-width=auto]
\hphantom{-}1 & \hphantom{-}0 \\
\hphantom{-}0 & C' 
\end{pNiceMatrix}
}
&
=
\frac{C}{2}
\underbrace
{
\begin{pNiceMatrix}[r]
1 & 0 \\
0 & 1 
\end{pNiceMatrix}
}
_
{\mathbf{I}}
+
\left(
1 - \frac{C}{2}
\right)
\underbrace
{
\setlength{\arraycolsep}{2pt}
\begin{pNiceMatrix}[r]
1 &  0 \\
0 & -1 
\end{pNiceMatrix}
}
_
{\mathbf{Z}}
\end{aligned}
\label{equation : block matrix decomposition : Robin boundary condition}
\\
\begin{aligned}
\mathbf{B}_{0}
&
=
\underbrace
{
\diag(1,\ldots,1,\frac{D}{2})_{N/2}
\otimes
\mathbf{I}
}
_
{\mathbf{D}_{0,\mathbf{I}}}
+
\underbrace
{
\diag(0,\ldots,0,1-\frac{D}{2})_{N/2}
\otimes
\mathbf{Z}
}
_
{\mathbf{D}_{0,\mathbf{Z}}}
-
\underbrace
{
\diag(1,\ldots,1,1)_{N/2}
\otimes
\mathbf{X}
}
_
{\mathbf{D}_{0,\mathbf{X}}}
\\
\mathbf{B}_{1}
&
=
\underbrace
{
\diag(1,\ldots,1,\frac{C}{2})_{N/2}
\otimes
\mathbf{I}
}
_
{\mathbf{D}_{1,\mathbf{I}}}
+
\underbrace
{
\diag(0,\ldots,0,1-\frac{C}{2})_{N/2}
\otimes
\mathbf{Z}
}
_
{\mathbf{D}_{1,\mathbf{Z}}}
-
\underbrace
{
\diag(1,\ldots,1,0)_{N/2}
\otimes
\mathbf{X}
}
_
{\mathbf{D}_{1,\mathbf{X}}}
\end{aligned}
\label{equation : block-diagonal matrix decomposition : Robin boundary condition}
\end{gather}

\begin{restatable}[Block-Diagonalization, Robin]{theorem}{theorem : block-diagonalization : Robin boundary condition}
\label{theorem : block-diagonalization : Robin boundary condition}
\begin{equation}
\mathbf{L}_R
=
\sum_{c \in \{0,1\}}
\sum_{\hat{\sigma} \in \{\mathbf{I},\mathbf{X},\mathbf{Z}\}}
\chi_{\hat{\sigma}}
\cdot
\mathbf{L}_{c, \hat{\sigma}}
\label{equation : block-diagonalization : Robin boundary condition}
\end{equation}
where
\begin{align}
\chi_{\hat{\sigma}}
&
:=
\begin{cases}
\hphantom{-}1, 
& 
\hat{\sigma} 
\in 
\{\mathbf{I},\mathbf{Z}\}
\\
-1, 
& 
\hat{\sigma} = \mathbf{X}
\end{cases}
\label{equation : coefficient : Robin boundary condition}
\\
\mathbf{L}_{c, \hat{\sigma}}
&
:=
\mathbf{P}_c
\cdot
\mathbf{D}_{c,\hat{\sigma}}
\cdot
\mathbf{P}_c^\dagger
\label{equation : subsystem : Robin boundary condition}
\\
\mathbf{D}_{c,\hat{\sigma}}
&
:=
\begin{cases}
\diag(1,\ldots,1,D/2) 
\otimes 
\hat{\sigma},
&
(c, \hat{\sigma}) = (0,\mathbf{I})
\\
\diag(1,\ldots,1,C/2)
\otimes 
\hat{\sigma},
&
(c, \hat{\sigma}) = (1,\mathbf{I})
\\
\diag(0,\ldots,0,1-D/2) 
\otimes 
\hat{\sigma},
&
(c, \hat{\sigma}) = (0,\mathbf{Z})
\\
\diag(0,\ldots,0,1-C/2)
\otimes 
\hat{\sigma},
&
(c, \hat{\sigma}) = (1,\mathbf{Z})
\\
\diag(1,\ldots,1,1)
\otimes 
\hat{\sigma},
&
(c, \hat{\sigma}) = (0,\mathbf{X})
\\
\diag(1,\ldots,1,0)
\otimes 
\hat{\sigma},
&
(c, \hat{\sigma}) = (1,\mathbf{X})
\end{cases}
\label{equation : block-diagonal : Robin boundary condition}
\end{align}
\end{restatable}

\subsection{Higher Dimensions}
\label{subsection : block-diagonalization : higher dimensions}

For higher dimensional problems, $\mathcal{L}$ is strictly speaking not block-diagonalizable by \cref{definition : block-diagonalizable matrix}. However, $\mathbf{L}$ which is a dependency of $\mathcal{L}$ is block-diagonalizable, as in \cref{theorem : block-diagonalization : periodic boundary condition,theorem : block-diagonalization : Dirichlet boundary condition,theorem : block-diagonalization : Neumann boundary condition,theorem : block-diagonalization : Robin boundary condition}. Therefore, we substitute its block-diagonalization into \cref{equation : placeholder : system of linear equations : higher dimensions}, then into \cref{equation : system of linear equations : higher dimensions}, and called the resulting form a block-diagonalization of $\mathcal{L}$, by abuse of definition.

\section{Block-Encoding}
\label{section : block-encoding}

With $\mathbf{L}_p$, $\mathbf{L}_D$, $\mathbf{L}_N$, $\mathbf{L}_R$ and $\mathcal{L}$ in block-diagonalization forms as in \cref{theorem : block-diagonalization : periodic boundary condition,theorem : block-diagonalization : Dirichlet boundary condition,theorem : block-diagonalization : Neumann boundary condition,theorem : block-diagonalization : Robin boundary condition} and \cref{subsection : block-diagonalization : higher dimensions}, we are ready to construct their respective block-encoding unitary operators using these forms. Several variations of block-encoding definition exist in the literature. In this article, we define block-encoding as in \cref{definition : block-encoding}, where $\mathcal{L}(\mathcal{H}^{\otimes n})$ is a space of linear operator mapping a Hilbert space $\smash{\mathcal{H}^{\otimes n} := (\mathbb{C}^2)^{\otimes n}}$ onto itself, for some $n \in \mathbb{N}$. By isomorphism and abuse of notation, we say that all the aforementioned $N \times N$ matrices are also linear operators of $\smash{\mathcal{L}(\mathcal{H}^{\otimes n})}$, where $n := \log(N)$.

\begin{definition}[Block-Encoding]
\label{definition : block-encoding}
A unitary operator $\smash{\bar{\mathbf{L}} \in \mathcal{L}(\mathcal{H}^{\otimes n+m})}$ is a block-encoding of $\smash{\mathbf{L} \in \mathcal{L}(\mathcal{H}^{\otimes n})}$, if there exists $\eta \in \mathbb{R}^+$ such that
\begin{gather}
\mathbf{L}
=
\eta
\cdot
\mathbf{\Pi}
\cdot
\bar{\mathbf{L}}
\cdot
\mathbf{\Pi}^\dagger
\quad
\text{where}
\quad
\mathbf{\Pi}
:=
\bra{0}^{\otimes m}
\otimes
\mathbf{I}^{\otimes n}
\label{equation : block-encoding}
\end{gather}
More succinctly, we say that $\bar{\mathbf{L}}$ block-encode $\mathbf{L}/\eta$ as its upper-left block-matrix as in
\begin{gather}
\bar{\mathbf{L}}
:=
{
\setlength{\arraycolsep}{0pt}
\begin{pNiceMatrix}[first-row,first-col,columns-width=auto]
& 
\mathbf{\Pi}      
& 
\tilde{\mathbf{\Pi}}
\\
\mathbf{\Pi}^\dagger          
& 
\mathbf{L}/\eta 
& 
\cdot 
\\
\tilde{\mathbf{\Pi}}^\dagger
& 
\cdot             
& 
\cdot 
\end{pNiceMatrix}
}
\label{equation : block-encoding matrix}
\end{gather}
where $\smash{\tilde{\mathbf{\Pi}}}$ is a projection operator, which is orthogonal complement to $\mathbf{\Pi}$.
\end{definition}

In this work, all the block-encoding contain an exact subnormalized matrix $\mathbf{L}/\eta$, rather than an approximation by some precision error $\varepsilon$. Moreover, all the matrices are not only square, but also Hermitian and positive definite. And finally, they can all be decomposed as a linear combination of unitary (LCU) operators with efficient implementations, which become evident later. Therefore, the majority of block-encodings used in this article is via LCU technique as in \cite{gilyen2019quantum}.

This section is also divided into five subsections, which describe the block-encoding of $\mathbf{L}_p$ with periodic boundary condition in \cref{subsection : block-diagonalization : periodic boundary condition}, $\mathbf{L}_D$ with Dirichlet boundary condition in \cref{subsection : block-encoding : Dirichlet boundary condition}, $\mathbf{L}_N$ with Neumann boundary condition in \cref{subsection : block-encoding : Neumann boundary condition}, $\mathbf{L}_R$ with Robin boundary condition in \cref{subsection : block-encoding : Robin boundary condition}, and $\mathcal{L}$ from higher dimensional problems in \cref{subsection : block-encoding : higher dimensions}.

\subsection{Periodic}
\label{subsection : block-encoding : periodic boundary condition}

We begin by rewriting each linear operator $\mathbf{L}_{c,\hat{\sigma}}$ from \cref{equation : subsystem : periodic boundary condition}, in terms of elementary quantum gates and circuits, as in \cref{equation : rewriting : periodic boundary condition}, where $\mathbf{ADD1}$ is a modulo-$2^n$ incrementation operation, which is one of the quantum arithmetic operations.
\begin{gather}
\begin{gathered}
\mathbf{L}_{0, \mathbf{I}}
=
\mathbf{L}_{1, \mathbf{I}}
=
\mathbf{I}^{\otimes n},
\quad
\mathbf{L}_{0, \mathbf{X}}
=
\mathbf{I}^{\otimes n-1}
\otimes
\mathbf{X}
\\
\mathbf{L}_{1, \mathbf{X}}
=
\mathbf{ADD1}
\cdot
\left(
\mathbf{I}^{\otimes n-1}
\otimes
\mathbf{X}
\right)
\cdot
\mathbf{ADD1}^\dagger
\\
\mathbf{ADD1}
:=
\sum_{i = 0}^{2^n - 1}
\ketbra{i + 1 \bmod 2^n}{i}
\end{gathered}
\label{equation : rewriting : periodic boundary condition}
\end{gather}

Since each $\mathbf{L}_{c,\hat{\sigma}}$ is a unitary operator, we can directly represent it using quantum circuit notation as in \cref{equation : quantum gates and circuits : periodic boundary condition}. Then, we construct a unitary operator $\smash{\bar{\mathbf{L}}_p}$ as in \cref{equation : linear combination of unitary operators : periodic boundary condition}, which block-encode $\mathbf{L}_p/4$ via LCU technique using $2$ ancilla qubits. \Cref{equation : linear combination of unitary operators simplified : periodic boundary condition} provides a simplified version of \cref{equation : linear combination of unitary operators : periodic boundary condition}, in terms of elementary quantum gates and quantum circuits.
\begin{gather}
\mathbf{L}_{0, \mathbf{I}}
:=
\mathbf{L}_{1, \mathbf{I}}
:=
\left\llbracket
\begin{ZX}[circuit,row sep={.075cm},column sep={.1cm}]
\ar[rr] 
\zxInputMulti{4}{n}
&
\zxGate[draw=none,fill=none]{}
&
\zxOutputMulti{4}{n} 
\\
&
\zxGate[draw=none,fill=none,rotate=90]{\cdots} & 
\\
\ar[rr] &
\zxGate[draw=none,fill=none]{} &
\\
\ar[rr] &
\zxGate[draw=none,fill=none]{} &
\end{ZX}
\right\rrbracket,
\
\mathbf{L}_{0, \mathbf{X}}
:=
\left\llbracket
\begin{ZX}[circuit,row sep={.075cm},column sep={.1cm}]
\ar[rr] 
\zxInputMulti{4}{n} 
&
\zxGate[draw=none,fill=none]{}
&
\zxOutputMulti{4}{n} 
\\
&
\zxGate[draw=none,fill=none,rotate=90]{\cdots} & 
\\
\ar[rr] &
\zxGate[draw=none,fill=none]{} & 
\\
\rar &
\zxGate{X} \rar &
\end{ZX}
\right\rrbracket,
\
\mathbf{L}_{1, \mathbf{X}}
:=
\left\llbracket
\begin{ZX}[circuit,row sep={.075cm},column sep={.1cm}]
\rar 
\zxInputMulti{4}{n} 
& 
\zxGateMulti{4}{1}{\rotatebox{-90}{$\mathbf{ADD1}^\dagger$}} 
\ar[rr] 
&
&
\zxGateMulti{4}{1}{\rotatebox{-90}{$\mathbf{ADD1}$}} 
\rar 
& 
\zxOutputMulti{4}{n} 
\\
&&
\zxGate[draw=none,fill=none,rotate=90]{\cdots} 
&&
\\
\rar 
& 
\ar[rr] 
&&
\rar 
&
\\
\rar 
& 
\rar 
&
\zxGate{X} \rar 
&
\rar 
&
\end{ZX}
\right\rrbracket
\label{equation : quantum gates and circuits : periodic boundary condition}
\\[.3cm]
\bar{\mathbf{L}}_p
:=
\left\llbracket
\begin{ZX}[circuit,row sep={.75cm,between origins},column sep={.2cm}]
\rar &
\zxGate{H} \rar &
\zxOCtrl{} \dar \rar &
\zxOCtrl{} \dar \rar &
\zxCtrl{} \dar \rar &
\zxCtrl{} \dar \rar &
\zxGate{H} \rar &
\\
\rar &
\zxGate{H} \rar &
\zxOCtrl{} \dar \rar &
\zxCtrl{} \dar \rar &
\zxOCtrl{} \dar \rar &
\zxCtrl{} \dar \rar &
\zxGate{H} \rar &
\\
\ar[Bn.={n},rr] &
&
\zxGate{+\mathbf{L}_{0, \mathbf{I}}} \rar[B] &
\zxGate{-\mathbf{L}_{0, \mathbf{X}}} \rar[B] &
\zxGate{+\mathbf{L}_{1, \mathbf{I}}} \rar[B] &
\zxGate{-\mathbf{L}_{1, \mathbf{X}}} \ar[Bn.={n},rr] &
&
\end{ZX}
\right\rrbracket
\label{equation : linear combination of unitary operators : periodic boundary condition}
\\[.3cm]
\bar{\mathbf{L}}_p
=
\left\llbracket
\begin{ZX}[circuit,row sep={.075cm},column sep={.1cm}]
\rar 
& 
\zxGate{H} 
\rar 
&
\zxGate{X} 
\rar 
& 
\zxCtrl{} 
\ar[ddd]
\rar 
&
\zxGate{X} 
\rar 
&
\zxCtrl{} 
\ar[ddd] 
\rar 
&
\zxGate{H} 
\rar 
&
\\
\rar 
& 
\zxGate{H} 
\ar[rr] 
&&
\zxCtrl{} 
\rar 
&
\zxGate{Z} 
\rar 
& 
\zxCtrl{} 
\rar 
&
\zxGate{H} 
\rar 
&
\\
\ar[rrrr] 
\zxInputMulti{4}{n} 
&&&&
\zxGateMulti{4}{1}{\rotatebox{-90}{$\mathbf{ADD1}^\dagger$}} 
\ar[rr] 
&&
\zxGateMulti{4}{1}{\rotatebox{-90}{$\mathbf{ADD1}$}} 
\rar 
& 
\zxOutputMulti{4}{n} 
\\
&
\zxGate[draw=none,fill=none,rotate=90]{\cdots} 
&
\zxGate[draw=none,fill=none,rotate=90]{\cdots} 
&
\zxGate[draw=none,fill=none,rotate=90]{\cdots} 
\ar[dd] 
&&
\zxGate[draw=none,fill=none,rotate=90]{\cdots} 
\ar[dd] 
&&
\\
\ar[rrrr] 
&&&&
\ar[rr] 
&&
\rar 
&
\\
\ar[rrr] 
&&&
\zxNot{} 
\rar 
&
\rar
&
\zxNot{} 
\rar 
&
\rar 
&
\end{ZX}
\right\rrbracket
\label{equation : linear combination of unitary operators simplified : periodic boundary condition}
\end{gather}

\subsection{Dirichlet}
\label{subsection : block-encoding : Dirichlet boundary condition}

We block-encode $\mathbf{L}_{D}$, similarly to the previous case, by first block-encoding each $\mathbf{L}_{c,\hat{\sigma}}$ term from \cref{equation : subsystem : Dirichlet boundary condition}. 

However, $\mathbf{L}_{1,\mathbf{X}}$ is not a unitary operator. Hence, an additional ancilla qubit is required to block-encode it via LCU technique, while $\mathbf{L}_{0,\mathbf{I}}, \mathbf{L}_{1,\mathbf{I}}, \mathbf{L}_{0,\mathbf{X}}$, which are unitary, can be block-encoded by concatenating a single ancilla qubit. The unitary operator which block-encodes each $\mathbf{L}_{c,\hat{\sigma}}$ is denoted $\smash{\bar{\mathbf{L}}_{c,\hat{\sigma}}}$ to differentiate between the two operators.
\begin{gather}
\begin{gathered}
\mathbf{L}_{0, \mathbf{I}}
=
\mathbf{L}_{1, \mathbf{I}}
=
\mathbf{I}^{\otimes n},
\quad
\mathbf{L}_{0, \mathbf{X}}
=
\mathbf{I}^{\otimes n-1}
\otimes
\mathbf{X}
\\
\mathbf{L}_{1, \mathbf{X}}
=
\mathbf{ADD1}
\cdot
\left(
\left(
\frac{1}{2}
\cdot
\mathbf{I}^{\otimes n-1}
+
\frac{1}{2}
\cdot
\mathbf{C}^{n-2}\mathbf{Z}
\right)
\otimes
\mathbf{X}
\right)
\cdot
\mathbf{ADD1}^\dagger
\end{gathered}
\label{equation : rewriting : Dirichlet boundary condition}
\\[.3cm]
\bar{\mathbf{L}}_{0, \mathbf{I}}
:=
\bar{\mathbf{L}}_{1, \mathbf{I}}
:=
\left\llbracket
\begin{ZX}[circuit,row sep={.075cm},column sep={.1cm}]
\ar[rr] &
\zxGate[draw=none,fill=none]{} & 
\\
\ar[rr] 
\zxInputMulti{4}{n} 
&
\zxGate[draw=none,fill=none]{}
&
\zxOutputMulti{4}{n} 
\\
&
\zxGate[draw=none,fill=none,rotate=90]{\cdots} & 
\\
\ar[rr] &
\zxGate[draw=none,fill=none]{} &
\\
\ar[rr] &
\zxGate[draw=none,fill=none]{} &
\end{ZX}
\right\rrbracket,
\
\bar{\mathbf{L}}_{0, \mathbf{X}}
:=
\left\llbracket
\begin{ZX}[circuit,row sep={.075cm},column sep={.1cm}]
\ar[rr] &
\zxGate[draw=none,fill=none]{} & 
\\
\ar[rr] 
\zxInputMulti{4}{n} &
\zxGate[draw=none,fill=none]{} &
\zxOutputMulti{4}{n} 
\\
&
\zxGate[draw=none,fill=none,rotate=90]{\cdots} & 
\\
\ar[rr] &
\zxGate[draw=none,fill=none]{} & 
\\
\rar &
\zxGate{X} \rar &
\end{ZX}
\right\rrbracket,
\
\bar{\mathbf{L}}_{1, \mathbf{X}}
:=
\left\llbracket
\begin{ZX}[circuit,row sep={.075cm},column sep={.1cm}]
\rar & 
\zxGate{H} \rar & 
\zxCtrl{} \ar[dd] \rar &
\zxGate{H} \rar &
\\
\rar \zxInputMulti{4}{n} & 
\zxGateMulti{4}{1}{\rotatebox{-90}{$\mathbf{ADD1}^\dagger$}} \rar &
\zxCtrl{} \rar &
\zxGateMulti{4}{1}{\rotatebox{-90}{$\mathbf{ADD1}$}} \rar & 
\zxOutputMulti{4}{n} 
\\
&
&
\zxGate[draw=none,fill=none,rotate=90]{\cdots} \dar &
&
\\
\rar & 
\rar &
\zxCtrl{} \rar &
\rar &
\\
\rar & 
\rar &
\zxGate{X} \rar &
\rar &
\end{ZX}
\right\rrbracket
\label{equation : quantum gates and circuits : Dirichlet boundary condition}
\\[.3cm]
\bar{\mathbf{L}}_D
:=
\left\llbracket
\begin{ZX}[circuit,row sep={.75cm,between origins},column sep={.2cm}]
\rar &
\zxGate{H} \rar &
\zxOCtrl{} \dar \rar &
\zxOCtrl{} \dar \rar &
\zxCtrl{} \dar \rar &
\zxCtrl{} \dar \rar &
\zxGate{H} \rar &
\\
\rar &
\zxGate{H} \rar &
\zxOCtrl{} \dar \rar &
\zxCtrl{} \dar \rar &
\zxOCtrl{} \dar \rar &
\zxCtrl{} \dar \rar &
\zxGate{H} \rar &
\\
\ar[Bn.={n+1},rr] &
&
\zxGate{+\bar{\mathbf{L}}_{0, \mathbf{I}}} \rar[B] &
\zxGate{-\bar{\mathbf{L}}_{0, \mathbf{X}}} \rar[B] &
\zxGate{+\bar{\mathbf{L}}_{1, \mathbf{I}}} \rar[B] &
\zxGate{-\bar{\mathbf{L}}_{1, \mathbf{X}}} \ar[Bn.={n+1},rr] &
&
\end{ZX}
\right\rrbracket
\label{equation : linear combination of unitary operators : Dirichlet boundary condition}
\\[.3cm]
\bar{\mathbf{L}}_D
=
\left\llbracket
\begin{ZX}[circuit,row sep={.075cm},column sep={.1cm}]
\rar & 
\zxGate{H} \rar &
\zxGate{X} \rar & 
\zxCtrl{} \dar \rar &
\zxGate{X} \rar &
\zxCtrl{} \ar[dddd] \rar &
\zxCtrl{} \ar[dddd] \rar &
\zxGate{H} \rar &
\\
\rar & 
\zxGate{H} \ar[rr] &
&
\zxCtrl{} \ar[ddd] \rar &
\zxGate{Z} \rar & 
\zxCtrl{} \rar &
\zxCtrl{} \rar &
\zxGate{H} \rar &
\\
\rar & 
\zxGate{H} \ar[rrrr] &
&
& 
&
\zxCtrl{} \ar[rr] &
&
\zxGate{H} \rar &
\\
\ar[rrrr] \zxInputMulti{4}{n} &
& 
&
&
\zxGateMulti{4}{1}{\rotatebox{-90}{$\mathbf{ADD1}^\dagger$}} \rar &
\zxCtrl{} \ar[rr] &
&
\zxGateMulti{4}{1}{\rotatebox{-90}{$\mathbf{ADD1}$}} \rar & 
\zxOutputMulti{4}{n} 
\\
&
\zxGate[draw=none,fill=none,rotate=90]{\cdots} &
\zxGate[draw=none,fill=none,rotate=90]{\cdots} &
\zxGate[draw=none,fill=none,rotate=90]{\cdots} \ar[dd] &
&
\zxGate[draw=none,fill=none,rotate=90]{\cdots} \dar &
\zxGate[draw=none,fill=none,rotate=90]{\cdots} \ar[dd] &
&
\\
\ar[rrrr] &
& 
&
&
\rar &
\zxCtrl{} \ar[rr] &
&
\rar &
\\
\ar[rrr] &
& 
&
\zxNot{} \rar &
\ar[rr] &
&
\zxNot{} \rar &
\rar &
\end{ZX}
\right\rrbracket
\label{equation : linear combination of unitary operators simplified : Dirichlet boundary condition}
\end{gather}

\subsection{Neumann}
\label{subsection : block-encoding : Neumann boundary condition}

We block-encode $\mathbf{L}_{N}$, similarly to the previous cases, by first block-encoding each $\mathbf{L}_{c,\hat{\sigma}}$ term from \cref{equation : subsystem : Neumann boundary condition}.
\begin{gather}
\begin{gathered}
\mathbf{L}_{0, \mathbf{I}}
=
\mathbf{I}^{\otimes n},
\quad
\mathbf{L}_{0, \mathbf{X}}
=
\mathbf{I}^{\otimes n-1}
\otimes
\mathbf{X}
\\
\mathbf{L}_{1, \mathbf{I}}
=
\mathbf{ADD1}
\cdot
\left(
\left(
\frac{1}{2}
\cdot
\mathbf{I}^{\otimes n-1}
+
\frac{1}{2}
\cdot
\mathbf{C}^{n-2}\mathbf{Z}
\right)
\otimes
\mathbf{I}
\right)
\cdot
\mathbf{ADD1}^\dagger
\\
\mathbf{L}_{1, \mathbf{X}}
=
\mathbf{ADD1}
\cdot
\left(
\left(
\frac{1}{2}
\cdot
\mathbf{I}^{\otimes n-1}
+
\frac{1}{2}
\cdot
\mathbf{C}^{n-2}\mathbf{Z}
\right)
\otimes
\mathbf{X}
\right)
\cdot
\mathbf{ADD1}^\dagger
\end{gathered}
\label{equation : rewriting : Neumann boundary condition}
\\[.3cm]
\begin{gathered}
\bar{\mathbf{L}}_{0, \mathbf{I}}
:=
\left\llbracket
\begin{ZX}[circuit,row sep={.075cm},column sep={.1cm}]
\ar[rr] &
\zxGate[draw=none,fill=none]{} & 
\\
\ar[rr]
\zxInputMulti{4}{n}
&
\zxGate[draw=none,fill=none]{}
&
\zxOutputMulti{4}{n} 
\\
&
\zxGate[draw=none,fill=none,rotate=90]{\cdots} & 
\\
\ar[rr] &
\zxGate[draw=none,fill=none]{} &
\\
\ar[rr] &
\zxGate[draw=none,fill=none]{} &
\end{ZX}
\right\rrbracket,
\
\bar{\mathbf{L}}_{0, \mathbf{X}}
:=
\left\llbracket
\begin{ZX}[circuit,row sep={.075cm},column sep={.1cm}]
\ar[rr] &
\zxGate[draw=none,fill=none]{} & 
\\
\ar[rr] 
\zxInputMulti{4}{n} &
\zxGate[draw=none,fill=none]{} &
\zxOutputMulti{4}{n} 
\\
&
\zxGate[draw=none,fill=none,rotate=90]{\cdots} & 
\\
\ar[rr] &
\zxGate[draw=none,fill=none]{} & 
\\
\rar &
\zxGate{X} \rar &
\end{ZX}
\right\rrbracket
\\[.3cm]
\bar{\mathbf{L}}_{1, \mathbf{I}}
:=
\left\llbracket
\begin{ZX}[circuit,row sep={.075cm},column sep={.1cm}]
\rar & 
\zxGate{H} \rar & 
\zxCtrl{} \ar[dd] \rar &
\zxGate{H} \rar &
\\
\rar \zxInputMulti{4}{n} & 
\zxGateMulti{4}{1}{\rotatebox{-90}{$\mathbf{ADD1}^\dagger$}} \rar &
\zxCtrl{} \rar &
\zxGateMulti{4}{1}{\rotatebox{-90}{$\mathbf{ADD1}$}} \rar & 
\zxOutputMulti{4}{n} 
\\
&
&
\zxGate[draw=none,fill=none,rotate=90]{\cdots} \dar &
&
\\
\rar & 
\rar &
\zxCtrl{} \rar &
\rar &
\\
\rar & 
\ar[rr] &&
\rar &
\end{ZX}
\right\rrbracket,
\
\bar{\mathbf{L}}_{1, \mathbf{X}}
:=
\left\llbracket
\begin{ZX}[circuit,row sep={.075cm},column sep={.1cm}]
\rar & 
\zxGate{H} \rar & 
\zxCtrl{} \ar[dd] \rar &
\zxGate{H} \rar &
\\
\rar \zxInputMulti{4}{n} & 
\zxGateMulti{4}{1}{\rotatebox{-90}{$\mathbf{ADD1}^\dagger$}} \rar &
\zxCtrl{} \rar &
\zxGateMulti{4}{1}{\rotatebox{-90}{$\mathbf{ADD1}$}} \rar & 
\zxOutputMulti{4}{n} 
\\
&
&
\zxGate[draw=none,fill=none,rotate=90]{\cdots} \dar &
&
\\
\rar & 
\rar &
\zxCtrl{} \rar &
\rar &
\\
\rar & 
\rar &
\zxGate{X} \rar &
\rar &
\end{ZX}
\right\rrbracket
\end{gathered}
\label{equation : quantum gates and circuits : Neumann boundary condition}
\\[.3cm]
\bar{\mathbf{L}}_N
:=
\left\llbracket
\begin{ZX}[circuit,row sep={.75cm,between origins},column sep={.2cm}]
\rar &
\zxGate{H} \rar &
\zxOCtrl{} \dar \rar &
\zxOCtrl{} \dar \rar &
\zxCtrl{} \dar \rar &
\zxCtrl{} \dar \rar &
\zxGate{H} \rar &
\\
\rar &
\zxGate{H} \rar &
\zxOCtrl{} \dar \rar &
\zxCtrl{} \dar \rar &
\zxOCtrl{} \dar \rar &
\zxCtrl{} \dar \rar &
\zxGate{H} \rar &
\\
\ar[Bn.={n+1},rr] &
&
\zxGate{+\bar{\mathbf{L}}_{0, \mathbf{I}}} \rar[B] &
\zxGate{-\bar{\mathbf{L}}_{0, \mathbf{X}}} \rar[B] &
\zxGate{+\bar{\mathbf{L}}_{1, \mathbf{I}}} \rar[B] &
\zxGate{-\bar{\mathbf{L}}_{1, \mathbf{X}}} \ar[Bn.={n+1},rr] &
&
\end{ZX}
\right\rrbracket
\label{equation : linear combination of unitary operators : Neumann boundary condition}
\\[.3cm]
\bar{\mathbf{L}}_N
=
\left\llbracket
\begin{ZX}[circuit,row sep={.075cm},column sep={.1cm}]
\rar & 
\zxGate{H} \rar &
\zxGate{X} \rar & 
\zxCtrl{} \dar \rar &
\zxGate{X} \rar &
\zxCtrl{} \ar[dddd] \rar &
\zxCtrl{} \ar[dddd] \rar &
\zxGate{H} \rar &
\\
\rar & 
\zxGate{H} \ar[rr] &
&
\zxCtrl{} \ar[ddd] \rar &
\zxGate{Z} \ar[rr] &&
\zxCtrl{} \rar &
\zxGate{H} \rar &
\\
\rar & 
\zxGate{H} \ar[rrrr] &
&
& 
&
\zxCtrl{} \ar[rr] &
&
\zxGate{H} \rar &
\\
\ar[rrrr] \zxInputMulti{4}{n} &
& 
&
&
\zxGateMulti{4}{1}{\rotatebox{-90}{$\mathbf{ADD1}^\dagger$}} \rar &
\zxCtrl{} \ar[rr] &
&
\zxGateMulti{4}{1}{\rotatebox{-90}{$\mathbf{ADD1}$}} \rar & 
\zxOutputMulti{4}{n} 
\\
&
\zxGate[draw=none,fill=none,rotate=90]{\cdots} &
\zxGate[draw=none,fill=none,rotate=90]{\cdots} &
\zxGate[draw=none,fill=none,rotate=90]{\cdots} \ar[dd] &
&
\zxGate[draw=none,fill=none,rotate=90]{\cdots} \dar &
\zxGate[draw=none,fill=none,rotate=90]{\cdots} \ar[dd] &
&
\\
\ar[rrrr] &
& 
&
&
\rar &
\zxCtrl{} \ar[rr] &
&
\rar &
\\
\ar[rrr] &
& 
&
\zxNot{} \rar &
\ar[rr] &
&
\zxNot{} \rar &
\rar &
\end{ZX}
\right\rrbracket
\label{equation : linear combination of unitary operators simplified : Neumann boundary condition}
\end{gather}

\subsection{Robin}
\label{subsection : block-encoding : Robin boundary condition}

We block-encode $\mathbf{L}_{R}$, similarly to the previous cases, by first block-encoding each $\mathbf{L}_{c,\hat{\sigma}}$ term from \cref{equation : subsystem : Robin boundary condition}.

One caveat regarding the Robin boundary condition in comparison to previous cases, is that the number of $\mathbf{L}_{c,\hat{\sigma}}$ terms is $6$ rather than $4$, which is not a power of $2$. To circumvent this issue, we construct a block-encoding of a zero operator $\mathbf{0}$, denoted $\bar{\mathbf{0}}$ in \cref{equation : quantum gates and circuits : Robin boundary condition}, and think of $\mathbf{L}_R$ as a linear combination of $8$ terms: six $\mathbf{L}_{c,\hat{\sigma}}$ and two $\mathbf{0}$ terms of coefficient $1$.

Another caveat is that the block-encoding quantum circuits of $\mathbf{L}_{0,\mathbf{I}}$, $\mathbf{L}_{0,\mathbf{Z}}$ and $\mathbf{L}_{1,\mathbf{I}}$, $\mathbf{L}_{1,\mathbf{Z}}$ can change, depending on the values of $D$ and $C$, respectively. In this work, we choose $D,C \in [0,2)$ so that the coefficients $1 \pm D/2$ and $1 \pm C/2$ are positive. This leads to block-encodings with subnormalization constant $1$ for $\mathbf{L}_{0,\mathbf{I}}$, $\mathbf{L}_{1,\mathbf{I}}$, and $1-D/2$ and $1-C/2$ for $\mathbf{L}_{0,\mathbf{Z}}$ and $\mathbf{L}_{1,\mathbf{Z}}$, respectively. Conventionally, to further block-encode $\mathbf{L}_{0,\mathbf{Z}}$ and $\mathbf{L}_{1,\mathbf{Z}}$ as a linear combination terms of $\mathbf{L}_R$ with coefficient $1$ as in \cref{equation : block-diagonalization : Robin boundary condition}, we need a quantum state preparation subroutine that encode $1 - D/2$ and $1 - C/2$. In this work, we keep the previous uniform quantum state preparation, and instead use controlled-$\mathbf{R}_Y$ gates to encode $1 - D/2$ and $1 - C/2$ on another ancilla, as in \cref{equation : linear combination of unitary operators : Robin boundary condition}. This results in a quantum circuit $\bar{\mathbf{L}}_R$ block-encoding $\mathbf{L}_R/8$. A similar block-encoding can be constructed for other $D,C$ values, using at most a constant number of additional gates.
\begin{gather}
\begin{gathered}
\mathbf{L}_{0, \mathbf{I}}
=
\left(
\frac{1}{2}
(1 + \frac{D}{2})
\cdot
\mathbf{I}^{\otimes n-1}
+
\frac{1}{2}
(1 - \frac{D}{2})
\cdot
\mathbf{C}^{n-2}\mathbf{Z}
\right)
\otimes
\mathbf{I}
\\
\mathbf{L}_{0, \mathbf{Z}}
=
\left(
\frac{1}{2}
(
1 - \frac{D}{2}
)
\cdot
\mathbf{I}^{\otimes n-1}
-
\frac{1}{2}
(
1 - \frac{D}{2}
)
\cdot
\mathbf{C}^{n-2}\mathbf{Z}
\right)
\otimes
\mathbf{Z}
\\
\mathbf{L}_{0, \mathbf{X}}
=
\mathbf{I}^{\otimes n-1}
\otimes
\mathbf{X}
\\
\mathbf{L}_{1, \mathbf{I}}
=
\mathbf{ADD1}
\cdot
\left(
\left(
\frac{1}{2}
(
1 + \frac{C}{2}
)
\cdot
\mathbf{I}^{\otimes n-1}
+
\frac{1}{2}
(
1 - \frac{C}{2}
)
\cdot
\mathbf{C}^{n-2}\mathbf{Z}
\right)
\otimes
\mathbf{I}
\right)
\cdot
\mathbf{ADD1}^\dagger
\\
\mathbf{L}_{1, \mathbf{Z}}
=
\mathbf{ADD1}
\cdot
\left(
\left(
\frac{1}{2}
(
1 - \frac{C}{2}
)
\cdot
\mathbf{I}^{\otimes n-1}
-
\frac{1}{2}
(
1 - \frac{C}{2}
)
\cdot
\mathbf{C}^{n-2}\mathbf{Z}
\right)
\otimes
\mathbf{Z}
\right)
\cdot
\mathbf{ADD1}^\dagger
\\
\mathbf{L}_{1, \mathbf{X}}
=
\mathbf{ADD1}
\cdot
\left(
\left(
\frac{1}{2}
\cdot
\mathbf{I}^{\otimes n-1}
+
\frac{1}{2}
\cdot
\mathbf{C}^{n-2}\mathbf{Z}
\right)
\otimes
\mathbf{X}
\right)
\cdot
\mathbf{ADD1}^\dagger
\end{gathered}
\label{equation : rewriting : Robin boundary condition}
\\[.3cm]
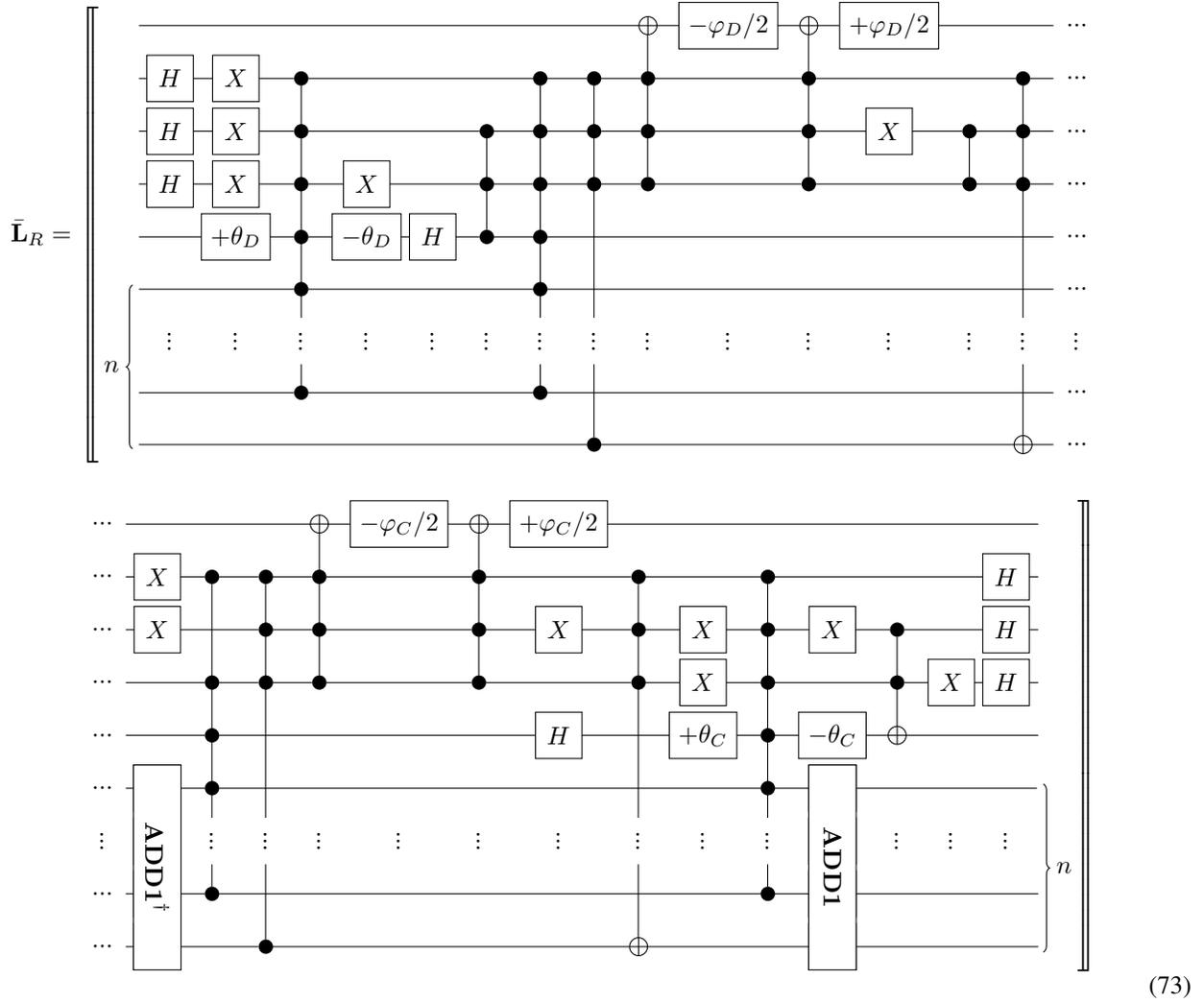
\begin{gathered}
\bar{\mathbf{L}}_{0, \mathbf{I}}
:=
\left\llbracket
\begin{ZX}[circuit,row sep={.075cm},column sep={.1cm}]
\ar[r] &
\zxGate{+\theta_{D}} \ar[r] &
\zxCtrl{} \ar[r] \ar[dd] &
\zxGate{-\theta_{D}} \ar[r] &
\\
\ar[rr] 
\zxInputMulti{4}{n} &
\zxGate[draw=none,fill=none]{} &
\zxCtrl{} \ar[rr] &&
\zxOutputMulti{4}{n} 
\\
&
\zxGate[draw=none,fill=none,rotate=90]{\cdots} &
\zxGate[draw=none,fill=none,rotate=90]{\cdots} \ar[d] &
\zxGate[draw=none,fill=none,rotate=90]{\cdots} &
\\
\ar[rr] &
\zxGate[draw=none,fill=none]{} &
\zxCtrl{} \ar[rr] &&
\\
\ar[rrrr] &
\zxGate[draw=none,fill=none]{} &&&
\end{ZX}
\right\rrbracket,
\
\bar{\mathbf{L}}_{1, \mathbf{I}}
:=
\left\llbracket
\begin{ZX}[circuit,row sep={.075cm},column sep={.1cm}]
\rar & 
\zxGate{+\theta_{C}} \ar[r] & 
\zxCtrl{} \ar[dd] \rar &
\zxGate{-\theta_{C}} \ar[r] &
\\
\rar \zxInputMulti{4}{n} & 
\zxGateMulti{4}{1}{\rotatebox{-90}{$\mathbf{ADD1}^\dagger$}} \ar[r] &
\zxCtrl{} \rar &
\zxGateMulti{4}{1}{\rotatebox{-90}{$\mathbf{ADD1}$}} \rar & 
\zxOutputMulti{4}{n} 
\\
&
&
\zxGate[draw=none,fill=none,rotate=90]{\cdots} \dar &
&
\\
\rar & 
\ar[r] &
\zxCtrl{} \rar &
\rar &
\\
\rar &
\ar[rr] &&
\ar[r] &
\end{ZX}
\right\rrbracket,
\\[.3cm]
\bar{\mathbf{L}}_{0, \mathbf{Z}}
:=
\left\llbracket
\begin{ZX}[circuit,row sep={.075cm},column sep={.1cm}]
\ar[r] &
\zxGate{H} \ar[r] &
\zxGate{Z} \ar[r] &
\zxCtrl{} \ar[r] \ar[dd] &
\zxGate{H} \ar[r] &
\\
\ar[rrr] 
\zxInputMulti{4}{n} &
\zxGate[draw=none,fill=none]{} &&
\zxCtrl{} \ar[rr] &&
\zxOutputMulti{4}{n} 
\\
&
\zxGate[draw=none,fill=none,rotate=90]{\cdots} &
\zxGate[draw=none,fill=none,rotate=90]{\cdots} &
\zxGate[draw=none,fill=none,rotate=90]{\cdots} \ar[d] &
\zxGate[draw=none,fill=none,rotate=90]{\cdots} &
\\
\ar[rrr] &
\zxGate[draw=none,fill=none]{} &&
\zxCtrl{} \ar[rr] &&
\\
\ar[rrr] &&&
\zxGate{Z} \ar[rr] &&
\end{ZX}
\right\rrbracket,
\
\bar{\mathbf{L}}_{1, \mathbf{Z}}
:=
\left\llbracket
\begin{ZX}[circuit,row sep={.075cm},column sep={.1cm}]
\rar & 
\zxGate{H} \rar &
\zxGate{Z} \ar[r] & 
\zxCtrl{} \ar[dd] \rar &
\zxGate{H} \rar &
\\
\rar \zxInputMulti{4}{n} & 
\zxGateMulti{4}{1}{\rotatebox{-90}{$\mathbf{ADD1}^\dagger$}} \ar[rr] &&
\zxCtrl{} \rar &
\zxGateMulti{4}{1}{\rotatebox{-90}{$\mathbf{ADD1}$}} \rar & 
\zxOutputMulti{4}{n} 
\\
&
&
\zxGate[draw=none,fill=none,rotate=90]{\cdots} &
\zxGate[draw=none,fill=none,rotate=90]{\cdots} \dar &
&
\\
\rar & 
\ar[rr] &&
\zxCtrl{} \rar &
\rar &
\\
\rar & 
\ar[rr] &&
\zxGate{Z} \ar[r] &
\rar &
\end{ZX}
\right\rrbracket,
\\[.3cm]
\bar{\mathbf{L}}_{0, \mathbf{X}}
:=
\left\llbracket
\begin{ZX}[circuit,row sep={.075cm},column sep={.1cm}]
\ar[rr] &
\zxGate[draw=none,fill=none]{} & 
\\
\ar[rr] \zxInputMulti{4}{n} &
\zxGate[draw=none,fill=none]{} &
\zxOutputMulti{4}{n} 
\\
&
\zxGate[draw=none,fill=none,rotate=90]{\cdots} & 
\\
\ar[rr] &
\zxGate[draw=none,fill=none]{} & 
\\
\rar &
\zxGate{X} \rar &
\end{ZX}
\right\rrbracket,
\
\bar{\mathbf{L}}_{1, \mathbf{X}}
:=
\left\llbracket
\begin{ZX}[circuit,row sep={.075cm},column sep={.1cm}]
\rar & 
\zxGate{H} \rar & 
\zxCtrl{} \ar[dd] \rar &
\zxGate{H} \rar &
\\
\rar \zxInputMulti{4}{n} & 
\zxGateMulti{4}{1}{\rotatebox{-90}{$\mathbf{ADD1}^\dagger$}} \rar &
\zxCtrl{} \rar &
\zxGateMulti{4}{1}{\rotatebox{-90}{$\mathbf{ADD1}$}} \rar & 
\zxOutputMulti{4}{n} 
\\
&
&
\zxGate[draw=none,fill=none,rotate=90]{\cdots} \dar &
&
\\
\rar & 
\rar &
\zxCtrl{} \rar &
\rar &
\\
\rar & 
\rar &
\zxGate{X} \rar &
\rar &
\end{ZX}
\right\rrbracket,
\
\bar{\mathbf{0}}
:=
\left\llbracket
\begin{ZX}[circuit,row sep={.075cm},column sep={.1cm}]
\rar &
\zxGate{X} \rar &
\\
\ar[rr] \zxInputMulti{4}{n} &
\zxGate[draw=none,fill=none]{} &
\zxOutputMulti{4}{n} 
\\
&
\zxGate[draw=none,fill=none,rotate=90]{\cdots} & 
\\
\ar[rr] && 
\\
\ar[rr] &
\zxGate[draw=none,fill=none]{} & 
\end{ZX}
\right\rrbracket
\end{gathered}
\label{equation : quantum gates and circuits : Robin boundary condition}
\shortintertext{where}
\begin{gathered}
\mathbf{R}_Y(\theta)
:=
\left\llbracket
\begin{ZX}[circuit,row sep={.075cm},column sep={.1cm}]
\ar[r] &
\zxGate{\theta} \ar[r] &
\end{ZX}
\right\rrbracket,
\quad
\theta_D
:=
2 
\cdot 
\arccos
\left(
\sqrt
{
\frac{1}{2}
(1 + \frac{D}{2})
}
\right)
\quad
\text{and}
\quad
\theta_C
:=
2 
\cdot 
\arccos
\left(
\sqrt
{
\frac{1}{2}
(1 + \frac{C}{2})
}
\right)
\end{gathered}
\label{equation : placeholder : quantum gates and circuits : Robin boundary condition}
\\
\bar{\mathbf{L}}_R
:=
\left\llbracket
\begin{ZX}[circuit,row sep={.75cm,between origins},column sep={.2cm}]
\ar[rrr]
&&&
\zxGate{\varphi_D} \ar[rrrr] \ar[d]
&&&&
\zxGate{\varphi_C} \ar[rrrr] \ar[d]
&&&&
\\
\rar &
\zxGate{H} \rar &
\zxOCtrl{} \dar \rar &
\zxOCtrl{} \dar \rar &
\zxOCtrl{} \dar \rar &
\zxOCtrl{} \dar \rar &
\zxCtrl{} \dar \rar &
\zxCtrl{} \dar \rar &
\zxCtrl{} \dar \rar &
\zxCtrl{} \dar \rar &
\zxGate{H} \rar &
\\
\rar &
\zxGate{H} \rar &
\zxOCtrl{} \dar \rar &
\zxOCtrl{} \dar \rar &
\zxCtrl{} \dar \rar &
\zxCtrl{} \dar \rar &
\zxOCtrl{} \dar \rar &
\zxOCtrl{} \dar \rar &
\zxCtrl{} \dar \rar &
\zxCtrl{} \dar \rar &
\zxGate{H} \rar &
\\
\rar &
\zxGate{H} \rar &
\zxOCtrl{} \dar \rar &
\zxCtrl{} \dar \rar &
\zxOCtrl{} \dar \rar &
\zxCtrl{} \dar \rar &
\zxOCtrl{} \dar \rar &
\zxCtrl{} \dar \rar &
\zxOCtrl{} \dar \rar &
\zxCtrl{} \dar \rar &
\zxGate{H} \rar &
\\
\ar[Bn.={n+1},rr] &
&
\zxGate{+\bar{\mathbf{L}}_{0, \mathbf{I}}} \rar[B] &
\zxGate{+\bar{\mathbf{L}}_{0, \mathbf{Z}}} \rar[B] &
\zxGate{+\bar{\mathbf{0}}} \rar[B] &
\zxGate{-\bar{\mathbf{L}}_{0, \mathbf{X}}} \rar[B] &
\zxGate{+\bar{\mathbf{L}}_{1, \mathbf{I}}} 
\rar[B] &
\zxGate{+\bar{\mathbf{L}}_{1, \mathbf{Z}}} \rar[B] &
\zxGate{+\bar{\mathbf{0}}}
\rar[B] &
\zxGate{-\bar{\mathbf{L}}_{1, \mathbf{X}}}
\ar[Bn.={n+1},rr] &&
\end{ZX}
\right\rrbracket
\label{equation : linear combination of unitary operators : Robin boundary condition}
\shortintertext{where}
\varphi_D
:=
2 \cdot \arccos\left( 1 - \frac{D}{2} \right)
\quad
\text{and}
\quad
\varphi_C
:=
2 \cdot \arccos\left( 1 - \frac{C}{2} \right)
\label{equation : placeholder : linear combination of unitary operators : Robin boundary condition}
\\[.5cm]
\begin{multlined}
\shoveright[.5cm]
{
\bar{\mathbf{L}}_R
=
\left\llbracket
\begin{ZX}[circuit,row sep={.075cm},column sep={.1cm}]
\ar[rrrrrrrrr]
&&&&&&&&&
\zxNot{} \ar[r] \ar[ddd]
&
\zxGate{-\varphi_D/2} \ar[r]
&
\zxNot{} \ar[r] \ar[ddd]
&
\zxGate{+\varphi_D/2} \ar[rrr]
&&&
\zxGate[draw=none,fill=none]{\cdots}
&
\\
\ar[r]
&
\zxGate{H} \ar[r]
&
\zxGate{X} \ar[r]
&
\zxCtrl{} \ar[rrrr] \ar[ddddd]
&&&&
\zxCtrl{} \ar[r] \ar[ddddd]
&
\zxCtrl{} \ar[r] \ar[ddddd]
&
\zxCtrl{} \ar[rr]
&&
\zxCtrl{} \ar[rrr]
&&&
\zxCtrl{} \ar[r] \ar[ddddd]
&
\zxGate[draw=none,fill=none]{\cdots}
&
\\
\ar[r]
&
\zxGate{H} \ar[r]
&
\zxGate{X} \ar[r]
&
\zxCtrl{} \ar[rrr]
&&&
\zxCtrl{} \ar[r] \ar[dd]
&
\zxCtrl{} \ar[r]
&
\zxCtrl{} \ar[r]
&
\zxCtrl{} \ar[rr]
&&
\zxCtrl{} \ar[r]
&
\zxGate{X} \ar[r]
&
\zxCtrl{} \ar[r] \ar[d]
&
\zxCtrl{} \ar[r]
&
\zxGate[draw=none,fill=none]{\cdots}
&
\\
\ar[r]
&
\zxGate{H} \ar[r]
&
\zxGate{X} \ar[r]
&
\zxCtrl{} \ar[r]
&
\zxGate{X} \ar[rr]
&&
\zxCtrl{} \ar[r]
&
\zxCtrl{} \ar[r]
&
\zxCtrl{} \ar[r]
&
\zxCtrl{} \ar[rr]
&&
\zxCtrl{} \ar[rr]
&&
\zxCtrl{} \ar[r]
&
\zxCtrl{} \ar[r]
&
\zxGate[draw=none,fill=none]{\cdots}
&
\\
\ar[rr]
&&
\zxGate{+\theta_D} \ar[r]
&
\zxCtrl{} \ar[r]
&
\zxGate{-\theta_D} \ar[r]
&
\zxGate{H} \ar[r]
&
\zxCtrl{} \ar[r]
&
\zxCtrl{} \ar[rrrrrrrr]
&&&&&&&&
\zxGate[draw=none,fill=none]{\cdots}
&
\\
\zxInputMulti{4}{n} 
\ar[rrr]
&&&
\zxCtrl{} \ar[rrrr]
&&&&
\zxCtrl{} \ar[rrrrrrrr]
&&&&&&&&
\zxGate[draw=none,fill=none]{\cdots}
&
\\
&
\zxGate[draw=none,fill=none,rotate=90]{\cdots} 
&
\zxGate[draw=none,fill=none,rotate=90]{\cdots} 
&
\zxGate[draw=none,fill=none,rotate=90]{\cdots}
\ar[d] 
&
\zxGate[draw=none,fill=none,rotate=90]{\cdots} 
&
\zxGate[draw=none,fill=none,rotate=90]{\cdots} 
&
\zxGate[draw=none,fill=none,rotate=90]{\cdots} 
&
\zxGate[draw=none,fill=none,rotate=90]{\cdots} 
\ar[d]
&
\zxGate[draw=none,fill=none,rotate=90]{\cdots} 
\ar[dd]
&
\zxGate[draw=none,fill=none,rotate=90]{\cdots} 
&
\zxGate[draw=none,fill=none,rotate=90]{\cdots} 
&
\zxGate[draw=none,fill=none,rotate=90]{\cdots} 
&
\zxGate[draw=none,fill=none,rotate=90]{\cdots} 
&
\zxGate[draw=none,fill=none,rotate=90]{\cdots} 
&
\zxGate[draw=none,fill=none,rotate=90]{\cdots} 
\ar[dd]
&
\zxGate[draw=none,fill=none,rotate=90]{\cdots}
&
\\
\ar[rrr]
&&&
\zxCtrl{} \ar[rrrr]
&&&&
\zxCtrl{} \ar[rrrrrrrr]
&&&&&&&&
\zxGate[draw=none,fill=none]{\cdots}
&
\\
\ar[rrrrrrrr]
&&&&&&&&
\zxCtrl{} \ar[rrrrrr]
&&&&&&
\zxNot{} \ar[r]
&
\zxGate[draw=none,fill=none]{\cdots}
&
\end{ZX}
\right.{}
}
\\[.3cm]
\shoveleft[.5cm]
{
\left.{}
\begin{ZX}[circuit,row sep={.075cm},column sep={.1cm}]
\zxGate[draw=none,fill=none]{\cdots}
\ar[rrrr]
&&&&
\zxNot{} \ar[r] \ar[ddd]
&
\zxGate{-\varphi_C/2} \ar[r]
&
\zxNot{} \ar[r] \ar[ddd]
&
\zxGate{+\varphi_C/2} 
\ar[rrrrrrrr]
&&&&&&&&
\\
\zxGate[draw=none,fill=none]{\cdots} \ar[r]
&
\zxGate{X} \ar[r]
&
\zxCtrl{} \ar[r] \ar[ddddd]
&
\zxCtrl{} \ar[r] \ar[ddddd]
&
\zxCtrl{} \ar[rr]
&&
\zxCtrl{} \ar[rr]
&&
\zxCtrl{} \ar[rr] \ar[ddddd]
&&
\zxCtrl{} \ar[rrrr] \ar[ddddd]
&&&&
\zxGate{H} \ar[r]
&
\\
\zxGate[draw=none,fill=none]{\cdots} \ar[r]
&
\zxGate{X} \ar[rr]
&&
\zxCtrl{} \ar[r]
&
\zxCtrl{} \ar[rr]
&&
\zxCtrl{} \ar[r]
&
\zxGate{X} \ar[r]
&
\zxCtrl{} \ar[r]
&
\zxGate{X} \ar[r]
&
\zxCtrl{} \ar[r]
&
\zxGate{X} \ar[r]
&
\zxCtrl{} \ar[rr] \ar[dd]
&&
\zxGate{H} \ar[r]
&
\\
\zxGate[draw=none,fill=none]{\cdots} \ar[rr]
&&
\zxCtrl{} \ar[r]
&
\zxCtrl{} \ar[r]
&
\zxCtrl{} \ar[rr]
&&
\zxCtrl{} \ar[rr]
&&
\zxCtrl{} \ar[r]
&
\zxGate{X} \ar[r]
&
\zxCtrl{} \ar[rr]
&&
\zxCtrl{} \ar[r]
&
\zxGate{X} \ar[r]
&
\zxGate{H} \ar[r]
&
\\
\zxGate[draw=none,fill=none]{\cdots} \ar[rr]
&&
\zxCtrl{} \ar[rrrrr]
&&&&&
\zxGate{H} \ar[rr]
&&
\zxGate{+\theta_C} \ar[r]
&
\zxCtrl{} \ar[r]
&
\zxGate{-\theta_C} \ar[r]
&
\zxNot{} \ar[rrr]
&&&
\\
\zxGate[draw=none,fill=none]{\cdots} \ar[r]
&
\zxGateMulti{4}{1}{\rotatebox{-90}{$\mathbf{ADD1}^\dagger$}} \ar[r]
&
\zxCtrl{} \ar[rrrrrrrr]
&&&&&&&&
\zxCtrl{} \ar[r]
&
\zxGateMulti{4}{1}{\rotatebox{-90}{$\mathbf{ADD1}$}} 
\ar[rrrr]
&&&&
\zxOutputMulti{4}{n}
\\
\zxGate[draw=none,fill=none,rotate=90]{\cdots}
&&
\zxGate[draw=none,fill=none,rotate=90]{\cdots} 
\ar[d]
&
\zxGate[draw=none,fill=none,rotate=90]{\cdots} 
\ar[dd]
&
\zxGate[draw=none,fill=none,rotate=90]{\cdots} 
&
\zxGate[draw=none,fill=none,rotate=90]{\cdots} 
&
\zxGate[draw=none,fill=none,rotate=90]{\cdots} 
&
\zxGate[draw=none,fill=none,rotate=90]{\cdots} 
&
\zxGate[draw=none,fill=none,rotate=90]{\cdots} 
\ar[dd]
&
\zxGate[draw=none,fill=none,rotate=90]{\cdots} 
&
\zxGate[draw=none,fill=none,rotate=90]{\cdots} 
\ar[d]
&&
\zxGate[draw=none,fill=none,rotate=90]{\cdots} 
&
\zxGate[draw=none,fill=none,rotate=90]{\cdots} 
&
\zxGate[draw=none,fill=none,rotate=90]{\cdots} 
&
\\
\zxGate[draw=none,fill=none]{\cdots} \ar[r]
&
\ar[r]
&
\zxCtrl{} \ar[rrrrrrrr]
&&&&&&&&
\zxCtrl{} \ar[r]
&
\ar[rrrr]
&&&&
\\
\zxGate[draw=none,fill=none]{\cdots} \ar[r]
&
\ar[rr]
&&
\zxCtrl{} \ar[rrrrr]
&&&&&
\zxNot{} \ar[rrr]
&&&
\ar[rrrr]
&&&&
\end{ZX}
\right\rrbracket
}
\end{multlined}
\label{equation : linear combination of unitary operators simplified : Robin boundary condition}
\end{gather}

\subsection{Higher Dimensions}
\label{subsection : block-encoding : higher dimensions}

For higher-dimensional cases, i.e., $d > 1$, we provide two alternative block-encoding schemes. The first scheme is given by a unitary operator $\bar{\mathcal{L}}$ as in \cref{equation : linear combination of unitary operators : higher dimensions}. It block-enocdes $\mathcal{L}/d$ via LCU technique using $\lceil\log(d)\rceil$ additional ancilla qubits. The original $\mathcal{O}(1)$ ancilla qubits used by block-encoding $\bar{\mathbf{L}}_1,\ldots,\bar{\mathbf{L}}_d$ can be shared together. In case that $\log(d) \not\in \mathbb{N}$, we need to construct additional block-encodings of a zero operator $\mathbf{0}$ for the remaining controls, as in the Robin case.
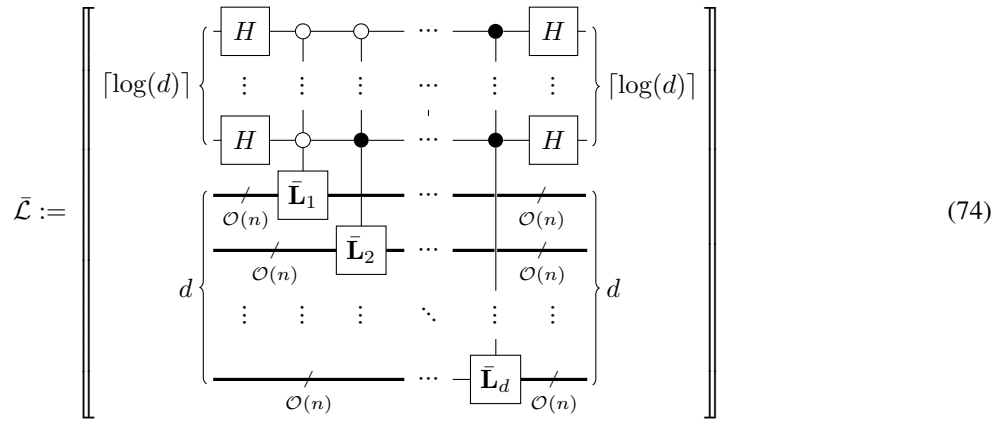
\begin{gather}
\bar{\mathcal{L}}
:=
\left\llbracket
\begin{ZX}[circuit,row sep={.075cm},column sep={.1cm}]
\rar
\zxInputMulti{3}{\lceil\log(d)\rceil} 
&
\zxGate{H}
\rar 
&
\zxOCtrl{} 
\dar 
\rar 
&
\zxOCtrl{} 
\dar 
\rar 
&
\zxGate[draw=none,fill=none]{\cdots} 
\rar 
&
\zxCtrl{} 
\dar 
\rar 
&
\zxGate{H} \rar 
&
\zxOutputMulti{3}{\lceil\log(d)\rceil}
\\
&
\zxGate[draw=none,fill=none,rotate=90]{\cdots} &
\zxGate[draw=none,fill=none,rotate=90]{\cdots} \dar &
\zxGate[draw=none,fill=none,rotate=90]{\cdots} \dar &
\zxGate[draw=none,fill=none]{\cdots} \dar &
\zxGate[draw=none,fill=none,rotate=90]{\cdots} \dar &
\zxGate[draw=none,fill=none,rotate=90]{\cdots} &
\\
\rar & 
\zxGate{H} \rar &
\zxOCtrl{} \dar \rar &
\zxCtrl{} \ar[dd] \rar &
\zxGate[draw=none,fill=none]{\cdots} \rar &
\zxCtrl{} \ar[ddd] \rar &
\zxGate{H} \rar &
\\
\zxInputMulti{4}{d}
\ar[Bn.={\mathcal{O}(n)},rr] &
&
\zxGate{\bar{\mathbf{L}}_1} \ar[rr,B] &&
\zxGate[draw=none,fill=none]{\cdots} 
\ar[B,r] &
\ar[rr,Bn.={\mathcal{O}(n)}] &&
\zxOutputMulti{4}{d}
\\
\ar[Bn.={\mathcal{O}(n)},rrr] &
&&
\zxGate{\bar{\mathbf{L}}_2} \ar[r,B]
&
\zxGate[draw=none,fill=none]{\cdots} 
\ar[B,r] &
\ar[rr,Bn.={\mathcal{O}(n)}] &&
\\
&
\zxGate[draw=none,fill=none,rotate=90]{\cdots} &
\zxGate[draw=none,fill=none,rotate=90]{\cdots} &
\zxGate[draw=none,fill=none,rotate=90]{\cdots} &
\zxGate[draw=none,fill=none,rotate=-45]{\cdots} &
\zxGate[draw=none,fill=none,rotate=90]{\cdots} \ar[d] &
\zxGate[draw=none,fill=none,rotate=90]{\cdots} &
\\
\ar[Bn.={\mathcal{O}(n)},rrrr] &
&&&
\zxGate[draw=none,fill=none]{\cdots} \rar &
\zxGate{\bar{\mathbf{L}}_d} \ar[rr,Bn.={\mathcal{O}(n)}] 
&&
\end{ZX}
\right\rrbracket
\label{equation : linear combination of unitary operators : higher dimensions}
\end{gather}

The second scheme is given by a unitary operator $\bar{\mathcal{L}}'$ as in \cref{equation : linear combination of unitary operators alternative : higher dimensions}. It block-encodes $\mathcal{L}/d$ via LCU technique using $\lceil\log(d)\rceil + d$ additional ancilla qubits. Unlike previous scheme, the original $\mathcal{O}(1)$ ancilla qubits from block-encodings $\bar{\mathbf{L}}_1,\ldots,\bar{\mathbf{L}}_d$ cannot be shared together. Therefore, the total number of ancillae is $\lceil\log(n)\rceil + \mathcal{O}(d)$. In case $\log(d) \not\in \mathbb{N}$, additional block-encodings of a zero operator $\mathbf{0}$ can be similarly added, as in the previous scheme and the Robin case.
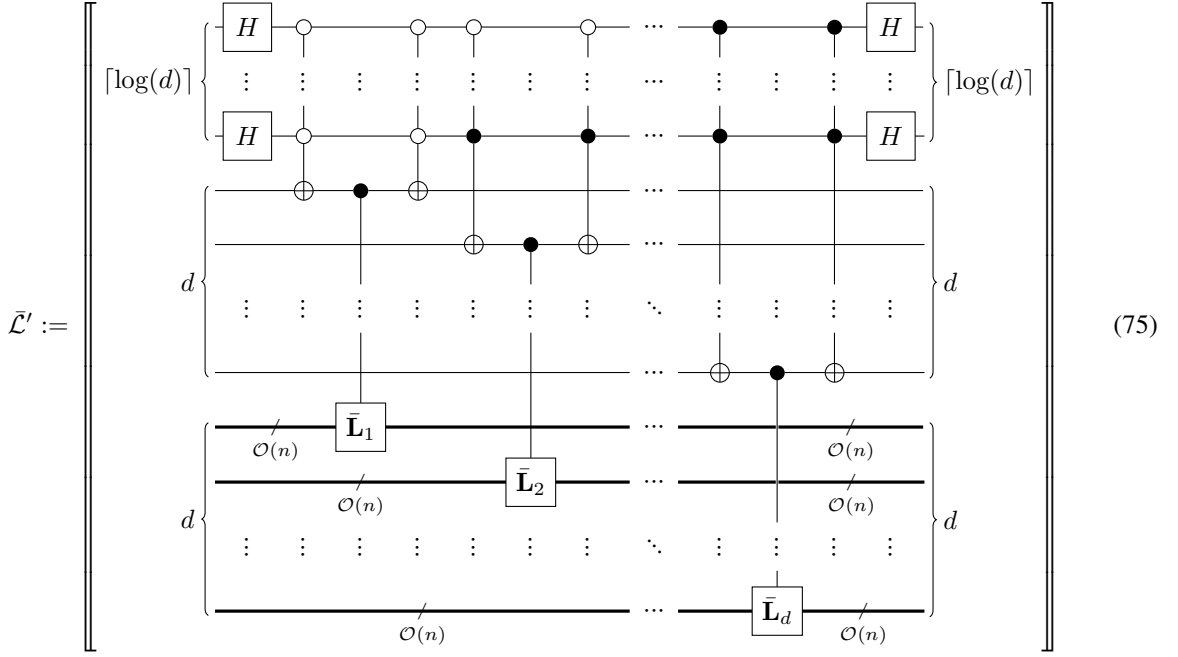
\begin{gather}
\bar{\mathcal{L}}'
:=
\left\llbracket
\begin{ZX}[circuit,row sep={.075cm},column sep={.1cm}]
\ar[r]
\zxInputMulti{3}{\lceil\log(d)\rceil} &
\zxGate{H} \ar[r] &
\zxOCtrl{} \ar[rr] \ar[d] &&
\zxOCtrl{} \ar[r] \ar[d] &
\zxOCtrl{} \ar[rr] \ar[d] &&
\zxOCtrl{} \ar[r] \ar[d] &
\zxGate[draw=none,fill=none]{\cdots} \ar[r] &
\zxCtrl{} \ar[rr] \ar[d] &&
\zxCtrl{} \ar[r] \ar[d] &
\zxGate{H} \ar[r] &
\zxOutputMulti{3}{\lceil\log(d)\rceil}
\\
&
\zxGate[draw=none,fill=none,rotate=90]{\cdots} &
\zxGate[draw=none,fill=none,rotate=90]{\cdots} \ar[d] &
\zxGate[draw=none,fill=none,rotate=90]{\cdots} &
\zxGate[draw=none,fill=none,rotate=90]{\cdots} \ar[d] &
\zxGate[draw=none,fill=none,rotate=90]{\cdots} \ar[d] &
\zxGate[draw=none,fill=none,rotate=90]{\cdots} &
\zxGate[draw=none,fill=none,rotate=90]{\cdots} \ar[d] &
\zxGate[draw=none,fill=none]{\cdots} &
\zxGate[draw=none,fill=none,rotate=90]{\cdots} \ar[d] &
\zxGate[draw=none,fill=none,rotate=90]{\cdots} &
\zxGate[draw=none,fill=none,rotate=90]{\cdots} \ar[d] &
\zxGate[draw=none,fill=none,rotate=90]{\cdots} &
\\
\ar[r] &
\zxGate{H} \ar[r] &
\zxOCtrl{} \ar[rr] \ar[d] &&
\zxOCtrl{} \ar[r] \ar[d] &
\zxCtrl{} \ar[rr] \ar[dd] &&
\zxCtrl{} \ar[r] \ar[dd] &
\zxGate[draw=none,fill=none]{\cdots} 
\ar[r] &
\zxCtrl{} \ar[rr] \ar[ddd] &&
\zxCtrl{} \ar[r] \ar[ddd] &
\zxGate{H} \ar[r] &
\\
\zxInputMulti{4}{d}
\ar[rr] &&
\zxNot{} \ar[r] &
\zxCtrl{} \ar[r] \ar[dd] &
\zxNot{} \ar[rrrr] &&&&
\zxGate[draw=none,fill=none]{\cdots} 
\ar[rrrrr] &&&&&
\zxOutputMulti{4}{d}
\\
\ar[rrrrr] &&&&&
\zxNot{} \ar[r] &
\zxCtrl{} \ar[r] \ar[d] &
\zxNot{} \ar[r] &
\zxGate[draw=none,fill=none]{\cdots}
\ar[rrrrr] &&&&&
\\
&
\zxGate[draw=none,fill=none,rotate=90]{\cdots} &
\zxGate[draw=none,fill=none,rotate=90]{\cdots} &
\zxGate[draw=none,fill=none,rotate=90]{\cdots} \ar[dd] &
\zxGate[draw=none,fill=none,rotate=90]{\cdots} &
\zxGate[draw=none,fill=none,rotate=90]{\cdots} &
\zxGate[draw=none,fill=none,rotate=90]{\cdots} \ar[ddd] &
\zxGate[draw=none,fill=none,rotate=90]{\cdots} &
\zxGate[draw=none,fill=none,rotate=-45]{\cdots} &
\zxGate[draw=none,fill=none,rotate=90]{\cdots} \ar[d] &
\zxGate[draw=none,fill=none,rotate=90]{\cdots} &
\zxGate[draw=none,fill=none,rotate=90]{\cdots} \ar[d] &
\zxGate[draw=none,fill=none,rotate=90]{\cdots} &
\\
\ar[rrrrrrrr] &&&&&&&&
\zxGate[draw=none,fill=none]{\cdots}
\ar[r] &
\zxNot{} \ar[r] &
\zxCtrl{} \ar[r] \ar[ddd] &
\zxNot{} \ar[rr] &&
\\
\zxInputMulti{4}{d}
\ar[Bn.={\mathcal{O}(n)},rrr] &&&
\zxGate{\bar{\mathbf{L}}_1}
\ar[B,rrrrr] &&&&&
\zxGate[draw=none,fill=none]{\cdots}
\ar[B,rr] &&
\ar[Bn.={\mathcal{O}(n)},rrr] &&&
\zxOutputMulti{4}{d}
\\
\ar[Bn.={\mathcal{O}(n)},rrrrrr] &&&&&&
\zxGate{\bar{\mathbf{L}}_2}
\ar[B,rr] &&
\zxGate[draw=none,fill=none]{\cdots}
\ar[B,rr] &&
\ar[Bn.={\mathcal{O}(n)},rrr] &&&
\\
&
\zxGate[draw=none,fill=none,rotate=90]{\cdots} &
\zxGate[draw=none,fill=none,rotate=90]{\cdots} &
\zxGate[draw=none,fill=none,rotate=90]{\cdots} &
\zxGate[draw=none,fill=none,rotate=90]{\cdots} &
\zxGate[draw=none,fill=none,rotate=90]{\cdots} &
\zxGate[draw=none,fill=none,rotate=90]{\cdots} &
\zxGate[draw=none,fill=none,rotate=90]{\cdots} &
\zxGate[draw=none,fill=none,rotate=-45]{\cdots} &
\zxGate[draw=none,fill=none,rotate=90]{\cdots} &
\zxGate[draw=none,fill=none,rotate=90]{\cdots} \ar[d] &
\zxGate[draw=none,fill=none,rotate=90]{\cdots} &
\zxGate[draw=none,fill=none,rotate=90]{\cdots} &
\\
\ar[Bn.={\mathcal{O}(n)},rrrrrrrr] &&&&&&&&
\zxGate[draw=none,fill=none]{\cdots}
\ar[B,rr] &&
\zxGate{\bar{\mathbf{L}}_d}
\ar[Bn.={\mathcal{O}(n)},rrr] &&&
\end{ZX}
\right\rrbracket
\label{equation : linear combination of unitary operators alternative : higher dimensions}
\end{gather}

\section{Analyses}
\label{section : analyses}

This section is divided into two parts. \Cref{subsection : analyses : block-encoding} analyses the computational resources, which include the number of ancilla qubits, gate count, and gate depth, required for the block-encoding quantum circuit of our block-diagonalization method. \Cref{subsection : analyses : boundary-value problems} analyses the time complexity for solving boundary-value problems, by utilizing the block-encoding quantum circuits and quantum linear solver algorithms to prepare the quantum state representation of the solution.

\subsection{Block-Encoding}
\label{subsection : analyses : block-encoding}

\Cref{subtable : quantum gates and circuits count} shows the computational resources required for the block-encoding quantum circuits $\smash{\bar{\mathbf{L}}_p}$, $\smash{\bar{\mathbf{L}}_D}$, $\smash{\bar{\mathbf{L}}_N}$ and $\smash{\bar{\mathbf{L}}_R}$, respectively. These resources include the number of ancilla qubits, quantum gates, and quantum circuits. For brevity, we count $\smash{\mathbf{ADD1}^{\dagger}}$ as the same quantum circuit as $\smash{\mathbf{ADD1}^{\vphantom{\dagger}}}$. \Cref{subtable : placeholder : quantum gates and circuits count} provides the description of each quantum gate. As observed, the computational complexities are dominated by quantum gate $\smash{\mathbf{C}^{\mathcal{O}(n)}\hat{\sigma}}$ and quantum circuit $\mathbf{ADD1}$, whose complexities depend on the number of qubit $n := \log(N)$ and their implementations in terms of simple quantum gates.

\begin{table}[h]
\centering
\begin{subtable}{\textwidth}
\centering
\begin{tabular}
{
@{}
l
>{\centering\arraybackslash}p{0.1\textwidth}
>{\centering\arraybackslash}p{0.05\textwidth}
>{\centering\arraybackslash}p{0.05\textwidth}
>{\centering\arraybackslash}p{0.05\textwidth}
>{\centering\arraybackslash}p{0.075\textwidth}
>{\centering\arraybackslash}p{0.075\textwidth}
>{\centering\arraybackslash}p{0.075\textwidth}
@{}
} 
\toprule
&
&
\multicolumn{3}{c}{\textbf{Single-Qubit Gates}} &
\multicolumn{2}{c}{\textbf{Multiple-Qubit Gates}} &
\\
\cmidrule(l){3-7}
&
Ancillae &
$\mathbf{H}$ &
$\hat{\sigma}$ &
$\mathbf{R}_{\hat{\sigma}}$ &
$\mathbf{C}^{\mathcal{O}(1)}\hat{\sigma}$ &
$\mathbf{C}^{\mathcal{O}(n)}\hat{\sigma}$ &
$\mathbf{ADD1}$ \\
\midrule
Periodic $\bar{\mathbf{L}}_p$ &
$2$ &
$4$ &
$3$ &
$0$ &
$2$ &
$0$ &
$2$ \\
\midrule
Dirichlet $\bar{\mathbf{L}}_D$ &
$3$ &
$6$ &
$3$ &
$0$ &
$2$ &
$1$ &
$2$ \\
\midrule
Neumann $\bar{\mathbf{L}}_N$ &
$3$ &
$6$ &
$3$ &
$0$ &
$2$ &
$1$ &
$2$ \\
\midrule
Robin $\bar{\mathbf{L}}_R$ &
$5$ &
$8$ &
$12$ &
$8$ &
$11$ &
$4$ &
$2$ \\
\bottomrule
\end{tabular}
\caption{Ancilla qubit, quantum gates, and quantum circuits count}
\label{subtable : quantum gates and circuits count}
\end{subtable}
\begin{subtable}{\textwidth}
\centering
\begin{tabular}{@{}ll@{}} 
\toprule
\textbf{Gate} &
\textbf{Description} \\
\midrule
$\mathbf{H}$ &
Hadamard Gate \\
\midrule
$\hat{\sigma}$ &
Pauli-$\hat{\sigma}$ Gate \\
\midrule
$\mathbf{R}_{\hat{\sigma}}$ &
Pauli-$\hat{\sigma}$ Rotation Gate \\
\midrule
$\mathbf{C}^{\mathcal{O}(1)}\hat{\sigma}$ &
$\mathcal{O}(1)$-Controlled Pauli-$\hat{\sigma}$ Gate \\
\midrule
$\mathbf{C}^{\mathcal{O}(n)}\hat{\sigma}$ &
$\mathcal{O}(n)$-Controlled Pauli-$\hat{\sigma}$ Gate \\
\bottomrule
\end{tabular}
\caption{Quantum gates description}
\label{subtable : placeholder : quantum gates and circuits count}
\end{subtable}
\caption{Computational resources required for the block-encoding quantum circuits $\smash{\bar{\mathbf{L}}_p}$, $\smash{\bar{\mathbf{L}}_D}$, $\smash{\bar{\mathbf{L}}_N}$ and $\smash{\bar{\mathbf{L}}_R}$}
\label{table : quantum gates and circuits count}
\end{table}

\Cref{subtable : quantum gates implementation,subtable : quantum circuits implementation} show the computational resource required to implement quantum gate $\mathbf{C}^{\mathcal{O}(n)\hat{\sigma}}$ and quantum circuit $\mathbf{ADD1}$, respectively. Because these implementations consist solely of classical reversible gates, they are given by the depth and count of Toffoli gate used; note that, the depth and count of smaller gates cannot exceed those of Toffoli gate. 

We present two implementations of $\mathbf{C}^{\mathcal{O}(n)}\hat{\sigma}$. The first implementation \cite{algassertUsingQuantum} has $\mathcal{O}(n)$ depth and count. This is achieved by borrowing one qubit which is not part of the quantum gate from the circuit, and thus does not require any additional ancilla qubit. Unfortunately, the second implementation \cite{gidney2018halving} does not provide their complexities. However, our analysis, denoted by * in \cref{subtable : quantum gates implementation}, suggests that they have $\mathcal{O}(\log(n))$ depth and $\mathcal{O}(n)$ count, using some $\mathcal{O}(n)$ additional ancillae. 

For $\mathbf{ADD1}$, we present $8$ alternative implementations. The first \cite{algassertUsingQuantum} once again has $\mathcal{O}(n)$ depth and count, and introduce no additional ancilla qubit by borrowing one existing qubit from the circuit. The rest \cite{draper2004logarithmic,takahashi2008fast,takahashi2009quantum,wang2023higher,wang2023reducing,wang2024optimal} are implementations of quantum addition circuit, which are specialized as a quantum modulo incrementation circuit instead. Unfortunately, except \cite{draper2004logarithmic}, the complexities are given for addition instead of modular incrementation. We denote this by * in \cref{subtable : quantum circuits implementation}. They have $\mathcal{O}(\log(n))$ depth, and $\mathcal{O}(n)$ or $\mathcal{O}(n\log(n))$ count, using $\mathcal{O}(n)$, $\mathcal{O}(n\log(n))$, or $\mathcal{O}(n/\log(n))$ ancillae. The best overall implementations are \cite{draper2004logarithmic,wang2023higher,wang2023reducing}, whose complexities are $\mathcal{O}(\log(n))$ depth, $\mathcal{O}(n)$ count, and $\mathcal{O}(n)$ ancillae.

Substituting $n$ by $\log(N)$, we can achieve a double-logarithmic depth $\mathcal{O}(\log(\log(N)))$ and a logarithmic $\mathcal{O}(\log(N))$ size implementation of our block-encoding quantum circuit, using implementations \cite{gidney2018halving,draper2004logarithmic,wang2023higher,wang2023reducing} and $\mathcal{O}(\log(N))$ ancilla qubits. Together with the original qubits, the total number of qubits is $c\log(N)+\mathcal{O}(1)$ for some $c \leq 12$. 

\begin{table}[h]
\centering
\begin{subtable}{\textwidth}
\centering
\begin{tabular}
{
@{}
l
>{\centering\arraybackslash}p{0.1\textwidth}
>{\centering\arraybackslash}p{0.1\textwidth}
>{\centering\arraybackslash}p{0.1\textwidth}
@{}
}
\toprule
\textbf{References} &
\textbf{Additional Ancillae} &
\textbf{Toffoli Depth} &
\textbf{Toffoli Count} \\
\midrule
\cite{algassertUsingQuantum} &
$0$ &
$\mathcal{O}(n)$ &
$\mathcal{O}(n)$ \\
\midrule
\cite{gidney2018halving}* &
$\mathcal{O}(n)$ &
$\mathcal{O}(\log(n))$ &
$\mathcal{O}(n)$ \\
\bottomrule
\end{tabular}
\caption{$\mathbf{C}^{\mathcal{O}(n)}\hat{\sigma}$}
\label{subtable : quantum gates implementation}
\end{subtable}
\begin{subtable}{\textwidth}
\centering
\begin{tabular}
{
@{}
l
>{\centering\arraybackslash}p{0.25\textwidth}
>{\centering\arraybackslash}p{0.25\textwidth}
>{\centering\arraybackslash}p{0.25\textwidth}
@{}
}
\toprule
\textbf{References} &
\textbf{Additional Ancillae} &
\textbf{Toffoli Depth} &
\textbf{Toffoli Count} \\
\midrule
\cite{algassertUsingQuantum} &
$0$ &
$\mathcal{O}(n)$ &
$\mathcal{O}(n)$ \\
\midrule
\cite{draper2004logarithmic} &
$2n-2\log(n)$ &
$2\log(n)+1$ &
$5n-6\log(n)-3$ \\
\midrule
\cite{takahashi2008fast}* &
$3n/\log(n) + n$ &
$30\log(n)$ &
$28n$ \\
\midrule
\cite{takahashi2009quantum}* &
$3n/\log(n) + n$ &
$18\log(n)$ &
$7n$ \\
\midrule
\cite{wang2023higher}* &
$6n-\log(n)+\mathcal{O}(1)$ &
$4\log(n)+\mathcal{O}(1)$ &
$8n-3\log(n)+\mathcal{O}(1)$ \\
\midrule
\cite{wang2023reducing}* &
$12n-6\log(n)+\mathcal{O}(1)$ &
$4\log(n)+\mathcal{O}(1)$ &
$13n-6\log(n)+\mathcal{O}(1)$ \\
\midrule
\cite{wang2024optimal}* &
$n\log(n)+n+\log(n)+2$ &
$2\log(n)+1$ &
$1.5n\log(n)-n-6\log(n)$ \\
\midrule
\cite{wang2024optimal}* &
$n\log(n)+n+\log(n)+2$ &
$\log(n)+1$ &
$0.5n\log(n)$ \\
\bottomrule
\end{tabular}
\caption{$\mathbf{ADD1}$}
\label{subtable : quantum circuits implementation}
\end{subtable}
\caption{Computational resources required for the implementation of $\mathbf{C}^{\mathcal{O}(n)}\hat{\sigma}$ and $\mathbf{ADD1}$}
\label{table : quantum gates and circuits implementation}
\end{table}

\Cref{table : quantum gates and circuits count : higher dimensions} shows the circuit depth, size, and number of ancillae of the block-encoding $\bar{\mathcal{L}}$ and $\bar{\mathcal{L}}'$ in higher dimensional cases. Both have roughly the same number of gate. However, $\bar{\mathcal{L}}'$ has lower depth and requires more ancilla qubits.

\begin{table}[h]
\centering
\begin{tabular}
{
@{}
l
>{\centering\arraybackslash}p{0.2\textwidth}
>{\centering\arraybackslash}p{0.2\textwidth}
>{\centering\arraybackslash}p{0.2\textwidth}
@{}
}
\toprule
&
\textbf{Total Ancillae} &
\textbf{Toffoli Depth} &
\textbf{Toffoli Count} \\
\midrule
Higher Dimensions $\bar{\mathcal{L}}$ &
$\mathcal{O}(\log(d))$ &
$\mathcal{O}(d\log(\log(N)))$ &
$\mathcal{O}(d\log(dN))$ \\
\midrule
Higher Dimensions $\bar{\mathcal{L}}'$ &
$\mathcal{O}(d+\log(dN))$ &
$\mathcal{O}(d+\log(\log(N)))$ &
$\mathcal{O}(d\log(dN))$ \\
\bottomrule
\\
\end{tabular}
\caption{Computational resources required for the block-encoding quantum circuit $\bar{\mathcal{L}}$ and $\bar{\mathcal{L}}'$}
\label{table : quantum gates and circuits count : higher dimensions}
\end{table}

\subsection{Boundary-Value Problems}
\label{subsection : analyses : boundary-value problems}

With our block-encoding quantum circuits, we are ready to solve our boundary-value problems. This is achieved by solving a system of linear equations $\mathcal{L}\ket{u}=\ket{f}$ via quantum linear solver algorithms \cite{harrow2009quantum,childs2017quantum,gilyen2019quantum,martyn2021grand,tong2021fast,algassertUsingQuantum,subacsi2019quantum,an2022quantum,lin2020optimal,costa2022optimal,jennings2023efficient,huang2019near,bravo2019variational,xu2021variational}. In this article, we use the quantum linear solver from \cite{costa2022optimal}, which at present has the best time complexity. It requires two quantum oracles: one block-encoding $\mathcal{L}$, and one preparing a quantum state encoding of $\smash{\ket{f}}$. It produces a quantum state encoding of $\mathcal{L}^{-1}\ket{f}$ to within $\epsilon$ error, using $\mathcal{O}(\kappa \log(1/\epsilon))$ calls to both oracles, where the condition number $\kappa := \smash{||\mathcal{L}^{-1}||}$ and $||\mathcal{L}||=1$.

\Cref{table : boundary-value problem time complexity} shows the time complexities of quantum linear solver \cite{costa2022optimal}, when applied to our boundary-value problems using our block-diagonalization block-encoding $\bar{\mathcal{L}}$ and $\bar{\mathcal{L}}'$. We compare our analyses against those of classical linear solvers, i.e., the conjugate gradient method \cite{saad2003iterative,vishnoi2013lx}, and quantum linear solvers using other matrix encoding methods \cite{dalzell2023quantum,childs2021high,kharazi2024explicit}. 

Basically, \cite[§7 solving differential equations]{dalzell2023quantum} refers to the conventional quantum oracle which computes the entries of the matrix given the row and column index, which we use together with quantum linear solver \cite{costa2022optimal}. While, \cite{childs2021high} refers to the adaptive finite difference method, with quantum linear solver \cite{childs2017quantum}. And, \cite{kharazi2024explicit} refers to the block-encoding technique using matrix decomposition similar to our block-diagonalization, with quantum linear solver \cite{costa2022optimal}. There are also several other quantum methods \cite{cao2013quantum,vazquez2022enhancing} as well; however, their techniques are similar to the aforementioned.

\Cref{subtable : boundary-value problem time complexity dependencies} shows the time complexities with dependencies in $\kappa$ the matrix's condition number, the number of grid point in one-dimension $N$, the dimension $d$, and the linear solver's solution error $\epsilon$. The primary observation is the dependency in $N$, where our method is exponentially better than existing quantum methods \cite{dalzell2023quantum,childs2021high,kharazi2024explicit} and superexponentially better than classical method \cite{saad2003iterative,vishnoi2013lx}. For other dependencies, the classical method is quadratically better than all quantum methods in $\kappa$, the quantum method \cite{childs2021high} is polynomially worse than others in $d$, and dependency in $\epsilon$ is the same for all.

\Cref{subtable : boundary-value problem time complexity dependencies alternative} shows the time complexities with dependencies in $d,\alpha,\delta,\epsilon$, where $\delta$ is the grid discretization error of the domain $\smash{[0,1]^d}$, and $\alpha=0.5$ represents a second-order approximation of the $3$-points central difference approximation scheme. They are derived by substituting $\kappa = \mathcal{O}(d)$ and $\smash{N = h^{-1} = \mathcal{O}((1/\delta)^{\alpha d})}$ into the complexities of \cref{subtable : boundary-value problem time complexity dependencies}. 

The primary dependency is $\delta$, where our methods are exponentially better than existing quantum methods \cite{dalzell2023quantum,childs2021high,kharazi2024explicit}, and superexponentially better than existing classical methods \cite{saad2003iterative,vishnoi2013lx}. The classical method is polynomially better than other methods by a small degree in $d$. Our methods are exponentially better in $\alpha$ and slightly better in $d$ than other quantum methods. The dependency in $\epsilon$ is once again the same for all. In the end-to-end analysis, we take $\delta = \mathcal{O}(\epsilon)$ \cite{dalzell2023quantum}, i.e., we assume that both errors are of the same order; in which case, our dependencies in $1/\epsilon$ are still polynomially better than \cite{childs2021high}, almost quadratically better than \cite{kharazi2024explicit}, and exponentially better than existing classical methods \cite{saad2003iterative,vishnoi2013lx}.

Despite the quantum advantages, there are several caveats in these analyses. First, we assume there exists a quantum state preparation of $\ket{f}$ with at worst $\mathcal{O}(\log(\log(N)))$ time complexity. As far as we know, except for some trivial cases, such quantum subroutine does not exist for general $\ket{f}$ \cite{dalzell2023quantum}. Secondly, we produce the quantum state representation of $\ket{u}$, rather than the classical state representation, which requires quantum state tomography with $\mathcal{O}(N)$ or $\smash{\mathcal{O}(\sqrt{N})}$ multiplicative overhead \cite{dalzell2023quantum}. Similarly, estimating a linear functional of $\ket{u}$ to within $\epsilon$ error, requires quantum amplitude estimation with $\mathcal{O}(||u||/\epsilon)$ multiplicative overhead \cite{dalzell2023quantum}. In either case, any gained quantum speedup diminishes.

\begin{table}[h]
\centering
\begin{subtable}{\textwidth}
\centering
\begin{tabular}
{
@{}
l
>{\centering\arraybackslash}p{0.45\textwidth}
@{}
}
\toprule
\textbf{References} &
\textbf{Dependencies in $\kappa,N,d,\epsilon$} \\
\midrule
Classical \cite{saad2003iterative,vishnoi2013lx} &
$\mathcal{O}(\sqrt{\kappa}d N^d\log(1/\epsilon))$ \\
\midrule
Quantum \cite{childs2021high,kharazi2024explicit,dalzell2023quantum} &
$\mathcal{O}(\kappa\poly(d,\log(N))\log(1/\epsilon))$ \\
\midrule
Block-Diagonalization $\bar{\mathcal{L}}$ &
$\mathcal{O}(\kappa d \log(\log(N))\log(1/\epsilon))$ \\
\midrule
Block-Diagonalization $\bar{\mathcal{L}}'$ &
$\mathcal{O}(\kappa\log(\log(N))\log(1/\epsilon) + \kappa d \log(1/\epsilon))$ \\
\bottomrule
\end{tabular}
\caption{$\kappa,N,d,\epsilon$}
\label{subtable : boundary-value problem time complexity dependencies}
\end{subtable}
\begin{subtable}{\textwidth}
\centering
\begin{tabular}
{
@{}
l
>{\centering\arraybackslash}p{0.45\textwidth}
@{}
}
\toprule
\textbf{References} &
\textbf{Dependencies in $d,\alpha,\delta,\epsilon$} \\
\midrule
Classical \cite{saad2003iterative,vishnoi2013lx} &
$\mathcal{O}(d^{1.5}(1/\delta)^{\alpha d}\log(1/\epsilon))$ \\
\midrule
Quantum \cite{childs2021high,kharazi2024explicit,dalzell2023quantum} &
$\mathcal{O}(d\poly(\alpha d,\log(1/\delta))\log(1/\epsilon))$ \\
\midrule
Block-Diagonalization $\bar{\mathcal{L}}$ &
$\mathcal{O}(d^2\log(\alpha d \log(1/\delta))\log(1/\epsilon))$ \\
\midrule
Block-Diagonalization $\bar{\mathcal{L}}'$ &
$\mathcal{O}(d\log(\alpha d\log(1/\delta))\log(1/\epsilon) + d^2\log(1/\epsilon))$ \\
\bottomrule
\end{tabular}
\caption{$d,\alpha,\delta,\epsilon$}
\label{subtable : boundary-value problem time complexity dependencies alternative}
\end{subtable}
\caption{Time complexity of boundary-value problems solved using classical and quantum linear solver algorithms}
\label{table : boundary-value problem time complexity}
\end{table}

\section{Discussion}
\label{section : discussion}

Solving differential equations is one of the primary applications of quantum computing; however, in spite of great advances towards a practical realization of an end-to-end application, several critical obstacles still remain. In this work, we address one of these obstacles, namely the encoding of the matrices representing the discretized forms of a Poisson partial differential equation within a quantum computer as quantum circuits. More specifically, this equation is given within the setting of boundary-value problems with hypercube domains and a number of different boundary conditions, i.e., periodic, Dirichlet, Neumann, and Robin, while the matrices are derived via finite difference method. The primary contribution of our work is the \emph{block-diagonalization} methodology. It provides a common decomposition form for all our matrices, whereby a common procedure for constructing quantum circuits, each of which encoding one of our matrices via the block-encoding technique. Additionally, a simplification of these quantum circuits in terms of elementary quantum gates and circuits can be trivially derived without the use of a sophisticated circuit optimization software, which shows the level of simplicities of these circuits. From the analyses and existing implementation techniques, we show that the computational resources required to construct each of these circuits are very efficient in terms of matrix size. Particularly, in conjunction with quantum linear solver algorithm, we can produce a quantum state solution of our boundary-value problems in time complexity double-logarithmic in the matrix size, using no more than a constant multiplicative overhead on the number of qubits and the number of gates. This represents an exponential and a superexponential improvement over existing quantum and classical methods, respectively. In terms of the inverse of solution error, it is at least polynomially and exponentially better than quantum and classical methods, respectively.

Our work represents an important step towards a practical realization of the uses of quantum computing for solving differential equations, by tackling one of the major obstacles in achieving this goal. We show that our methodology improves upon the existing methods in the literature, in terms of time complexity. In fact, as far as we know, this is the first method, which shows a double-logarithmic time complexity in the problem of encoding matrices from finite difference method. Several existing works are shown to have similar matrix decompositions or quantum circuits; however, our analyses are the first to show that a quantum circuit with a double-logarithmic depth in terms of matrix size, can be implemented with a small overhead in the number of ancilla qubits. Finally, this is but a first step for block-diagonalization methodology. We believe that it has the potential to be extended towards more complicated problems, or at least inspire a similar methodology to be developed, catering towards the specificities and complexities of each problem. These problems are not limited to solving differential equations; they can also include other quantum computing applications such as physics, chemistry, machine learning, continuous and combinatorial optimization.

The followings are the primary directions of future works, which are related to our block-diagonalization methodology.

\paragraph{Finite Difference Method}

The first obvious extension of block-diagonalization is to study its applications towards more complicated discretization schemes, or boundary-value problems, within the framework of finite difference method. For instance, it would be interesting to see whether it is possible to derive the block-diagonalization form for other discretization schemes, especially for higher-order schemes; or whether such derived forms, if found, also lead to a double-logarithmic depth quantum circuit implementation. Another interesting extension is for more complicated domains and domain boundaries, because the structure of their corresponding matrices are not as simple as those of hypercube domains, where we are able to derive the block-diagonalization form quite easily. And, the most important extensions are towards more complicated equations, including elliptic equations, hyperbolic equations, parabolic equations, and non-linear ordinary and partial differential equations, which are generally more difficult to solve.

\paragraph{General Numerical Methods}

Another line of extensions is towards other numerical methods for solving differential equations, such as finite element method, finite volume method, and spectral method. All these methods have a long history, and are developed to compensate for the weaknesses of finite difference method; for instance, for problems with domains of complex geometries. However, the majority of them results in a system of linear equations much like finite difference method, albeit one with a completely different structure. Then, the question is again whether it is also possible to derive a similar decomposition as block-diagonalization of these matrices, or whether such decomposition has a double-logarithmic depth implementation. Generally, these matrices are more complicated than those of finite difference method, especially their higher dimensional extensions. Together, they represent the majority of use cases in all the science and engineering fields; thus, their studies would greatly benefit a significant number of applications.

\paragraph{Solving Differential Equations}

Rather than extending the methodology towards other use cases, it is also imperative to study other quantum computing problems that can greatly aid in the practical realization of a complete end-to-end application of quantum algorithms towards solving differential equations. As pointed out several times throughout this article, the encoding of problem matrices is but one of the major obstacle towards this goal. For instance, to achieve an end-to-end double-logarithmic time complexity, we also need a quantum state preparation with a double-logarithmic depth in the matrix size, or a quantum linear solver algorithm with a double-logarithmic depth in the inverse of solution error. Similarly, the study of a quantum preconditioner would also greatly benefit the quantum linear solver algorithm. However, the most difficult challenge, we believe, is the measurement of the quantum state solution to obtain either the classical state solution or one of its properties, which has a linear depth in either the matrix size or the inverse of solution error. Together, they represent the primary obstacles towards a complete end-to-end practical realization.

\section*{Acknowledgement}
\label{section : acknowledgement}

The authors would like to thank David Danan, Pierre-Alain Boucard, and François Jouve, for their helpful comments, suggestions and discussions. All of whom help contribute towards improving the quality of both our work and article. This project is supported by the French government's aid in the framework of PIA (Programme d'Investissement d'Avenir) for Institut de Recherche Technologique SystemX, and from the PEPR integrated project EPiQ ANR-22-PETQ-0007 part of Plan France 2030.

\bibliographystyle{unsrt}
\bibliography{references}

\end{document}